\documentclass[8.5pt,twoside,twocolumn]{article}
\usepackage[super,sort&compress,comma]{natbib} 
\usepackage[pdftex]{graphicx}
\usepackage{times}
\usepackage{textcomp}
\usepackage[caption=false]{subfig}
\usepackage{amssymb,amsmath}
\graphicspath{{.},{Figs/}}

\usepackage{sectsty}
\usepackage{balance} 

\usepackage{graphicx} 
\usepackage{lastpage}
\usepackage[format=plain,justification=raggedright,singlelinecheck=false,font=small,labelfont=bf,labelsep=space]{caption} 
\usepackage{fancyhdr}
\begin{document}

\thispagestyle{plain}
\renewcommand{\thefootnote}{\fnsymbol{footnote}}
\renewcommand\footnoterule{\vspace*{1pt}%
\hrule width 3.4in height 0.4pt \vspace*{5pt}} 
\setcounter{secnumdepth}{5}

\makeatletter 
\def\subsubsection{\@startsection{subsubsection}{3}{10pt}{-1.25ex plus -1ex minus -.1ex}{0ex plus 0ex}{\normalsize\bf}} 
\def\paragraph{\@startsection{paragraph}{4}{10pt}{-1.25ex plus -1ex minus -.1ex}{0ex plus 0ex}{\normalsize\textit}} 
\renewcommand\@biblabel[1]{#1}            
\renewcommand\@makefntext[1]%
{\noindent\makebox[0pt][r]{\@thefnmark\,}#1}
\makeatother 
\renewcommand{\figurename}{\small{Fig.}~}
\sectionfont{\large}
\subsectionfont{\normalsize} 

\fancyfoot{}
\fancyfoot[RO]{\footnotesize{\sffamily{1--\pageref{LastPage} ~\textbar  \hspace{2pt}\thepage}}}
\fancyfoot[LE]{\footnotesize{\sffamily{\thepage~\textbar\hspace{3.45cm} 1--\pageref{LastPage}}}}
\fancyhead{}
\renewcommand{\headrulewidth}{1pt} 
\renewcommand{\footrulewidth}{1pt}
\setlength{\arrayrulewidth}{1pt}
\setlength{\columnsep}{6.5mm}
\setlength\bibsep{1pt}

\twocolumn[
  \begin{@twocolumnfalse}
\noindent\LARGE{\textbf{Correlation between surface topography and slippage:\\a Molecular Dynamics study}}
\vspace{0.6cm}

\noindent\large{\textbf{Nikita Tretyakov,$^{\ast}$ and
Marcus M\"{u}ller}}\vspace{0.5cm}

\noindent\textit{\small{\textbf{Received Xth XXXXXXXXXX 20XX, Accepted Xth XXXXXXXXX 20XX\newline
First published on the web Xth XXXXXXXXXX 200X}}}

\noindent \textbf{\small{DOI: 10.1039/b000000x}}
\vspace{0.6cm}

\noindent \normalsize{Using Molecular Dynamics simulations of a polymer liquid flowing past flat and patterned surfaces, we investigate the influence of corrugation, wettability and pressure on slippage and friction at the solid-liquid interface. For one-dimensional, shallow, rectangular grooves, we observe a gradual crossover between the  Wenzel state, where the liquid fills the grooves, and the Cassie state, where the corrugation supports the liquid and the grooves are filled with vapor. Using two independent flow set-ups, we characterize the near-surface flow by the slip length, $\delta$, and the position, $z_\textrm{h}$, at which viscous and frictional stresses are balanced according to Navier's partial slip boundary condition. This hydrodynamic boundary position depends on the pressure inside the channel and may be located above the corrugated surface. In the Cassie state, we observe that the edges of the corrugation contribute to the friction.}
\vspace{0.5cm}
 \end{@twocolumnfalse}
  ]

\footnotetext{\textit{~Institut f\"{u}r Theoretische Physik, Friedrich-Hund-Platz 1,
37077 G\"{o}ttingen, Germany. E-mail: Nikita.Tretyakov@theorie.physik.uni-goettingen.de}}

%
%
\section{Introduction}
\label{sec:intro}
%
%
Understanding the flow of liquids past solid surfaces has attracted abiding attention in micro- and nanofluidics.  While on macroscopic scales, the non-slip or stick boundary, which assumes the velocity of the fluid at the surface coincides with that of the surface,  is popular, this approximation may become invalid on smaller length scales pertinent to the operation of micro- and nanofluidic devices \cite{JO_BP_JR_2004, Rothstein_2010, MQM_MP_RRT_2011}. Controlling the friction at the interface between the solid substrate and the liquid is crucial in these applications, and different strategies have been pursued: Like in the case of wetting, one can control friction by (i) the direct, microscopic interactions between the solid and the liquid or (ii) the surface topography of the solid \cite{CCB_JLB_LB_EC_2003,TG_PA_2004,DK_ED_2006}. While the former is largely dictated by the chemistry of the solid and liquid, the latter alternative is expected to be a universal physical mechanism.

\begin{figure}[ht]
    \centering
    \includegraphics[height=4cm]{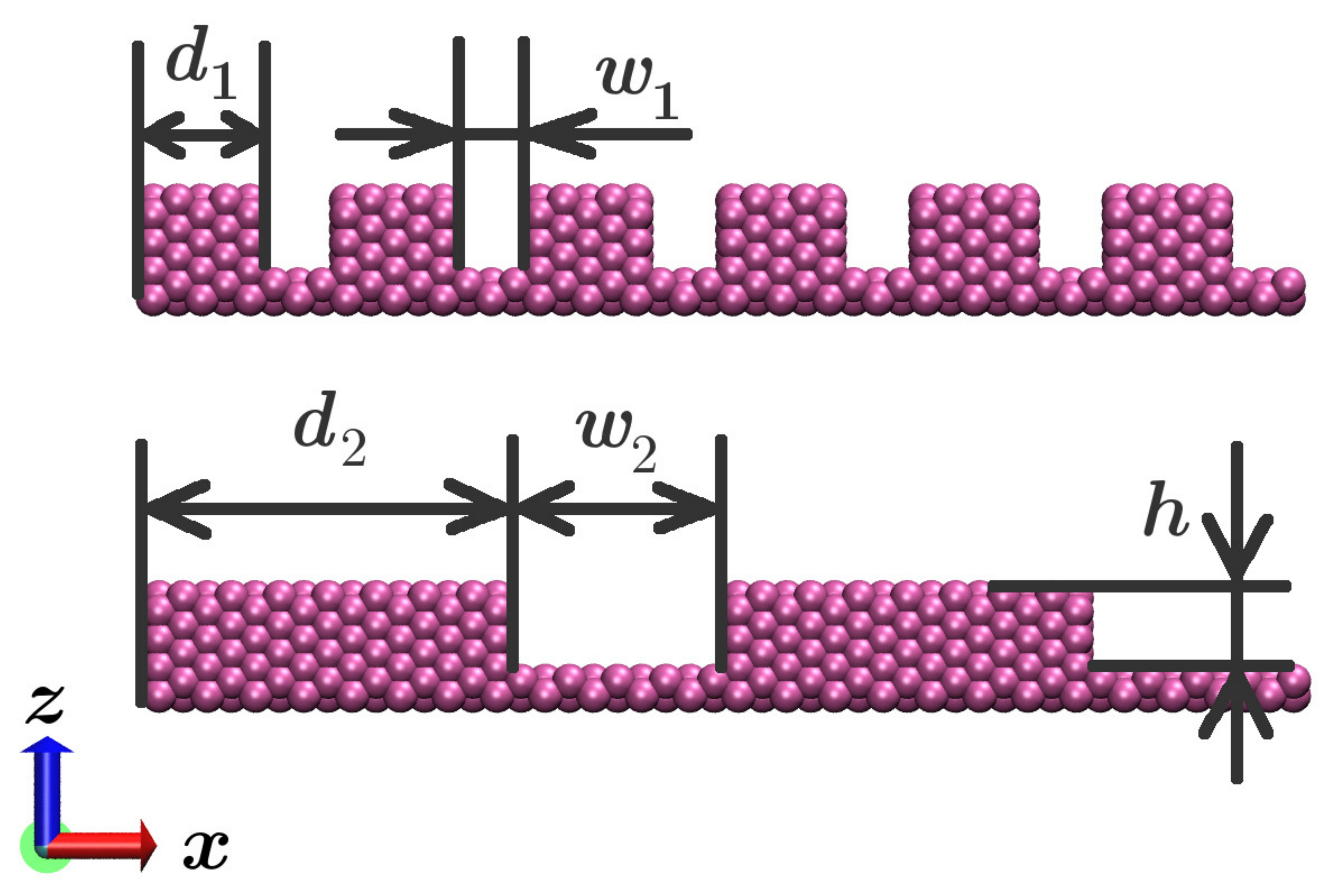}
    \caption{The geometries of the structured substrates. Patterning is present only in $x$-direction.}
    \label{fig:struct_sub}
\end{figure}

We consider a regularly corrugated surface topography as sketched in Fig.~\ref{fig:struct_sub}. A liquid on this structured substrate may either fill the cavities between corrugations (Wenzel state \cite{Wenzel_1936}) or build a straight liquid-vapor interface on the top of the corrugations (Cassie state \cite{AC_SB_1944} or fakir state). The macroscopic contact angle that a liquid drop makes with the solid substrate is dictated by the balance of surface tensions according to Young's equation \cite{Young}. Surface roughness tends to amplify wettability, i.e., if the liquid did not wet the planar solid substrate, the contact angle would be even larger on a corrugated solid \cite{Wenzel_1936, AC_SB_1944, Bico1999, Bico2003}.

The slippage of a liquid past a solid surface, in turn, is determined by a balance of viscous and frictional stresses at solid-liquid boundary~\cite{Navier_1823}. Modifying the topography of the surface, one simultaneously alters this balance and the surface tensions, therefore affecting both, slippage and wettability, respectively. A dramatical decrease of slippage takes place for a liquid in Wenzel state with respect to the corresponding flat substrate \cite{SG_YZ_HL_2003,CCB_JLB_LB_EC_2003,CCB_CB_EC_2004}. The effective position of equivalent no-slip wall was found between the top and bottom of the grooves \cite{CK_JH_OV_2010}.
In the case of the liquid in the Cassie (or fakir) state, in turn, the slip length increases and an enlargement by a factor of $2.5$ in comparison to flat substrates has been observed \cite{CCB_JLB_LB_EC_2003}. Semi-analytical models, assuming a no-slip condition at the solid-liquid interface and infinite slip at liquid-vapor interface, have been developed to calculate an effective slip~\cite{CCB_CB_EC_2004,OV_AB_2011}.

Both, in the description of wetting and slippage, the phenomenological approaches exploited the scale separation between the geometry of the substrate (i.e., the shape of the corrugation) and the length scale that determines the surface tension or friction of the corresponding planar substrate. This assumption is justified for macroscopic corrugations but when the spatial scale of the corrugation decreases, geometry and interactions do no longer decouple; in this limit the surface roughness is an intrinsic property of the substrate. In this manuscript, we use Molecular Dynamics simulations of a particle-based model, which duly account for the discreteness of the fluid and its thermal fluctuations, to investigate the onset of deviations from the macroscopic phenomenological behavior.

Keeping the fraction $\varphi$ of the surface covered by grooves constant but varying their spatial dimensions, we investigate the influence of the roughness onto dynamical properties of the liquid in the Cassie and Wenzel states. Between these two states of the liquid, we only observe a gradual crossover but no phase transition \cite{CCB_JLB_LB_EC_2003,AG_MC_SM_2012}, as the depth of the grooves is of the same order as the width of the liquid-vapor interface. We show that (i) the slippage at superhydrophobic substrates is significantly influenced by the additional friction at the edges of corrugations and (ii) that the hydrodynamic boundary position does not necessary coincide with a localization of the solid-liquid interface between the top and the bottom of the grooves. 

Our paper is organized as follows: In section~\ref{sec:model} we discuss the simulation model and techniques and  the macroscopic phenomenological models. In this section we also provide information about the equilibrium wetting behavior. The comparison of channels with patterned and flat walls is reported in sections~\ref{sec:Navier}-\ref{sec:comp-coex}. The study of flows in channels with patterned walls at different pressures is presented in section~\ref{sec:varpress}, and we briefly summarize our results in section~\ref{sec:discout}.
%
%
\section{Model and equilibrium wetting properties}
\label{sec:model}
%
%
\subsection{Coarse-grained polymer liquid and topographically structured substrate}
We represent the liquid by coarse-grained short chain molecules comprised of $N_\textrm{P}=10$ monomers \cite{Muller00f,Muller03c}. The excluded volume interaction and the cohesive attraction in the liquid are modeled by shifted Lennard-Jones (LJ) potential with energy scale $\epsilon$ and bead diameter $\sigma$. In the following, all energies are measured in units of $\epsilon$ and all length scales are given in units of $\sigma$. The LJ potential is truncated at $r_\textrm{cut}=2\times 2^{1/6} \sigma$. {Additionally,} the neighboring monomers belonging to one polymer chain are bonded by FENE potential \cite{Bird77, KK_GG_1990}. The virtue of using a polymer liquid is that the number density of the vapor phase is of the order of $\rho_{\rm{V}} \sim 10^{-5}\sigma^{-3}$, and one can neglect evaporation effects. Therefore the coexistence pressure is approximately zero. The density of the polymer liquid in coexistence with its vapor is $\rho_{\rm coex}=0.786 \sigma^{-3}$. 

\begin{figure}[ht]
    \centering
    \subfloat[]{\label{fig:cav_unfilled}\includegraphics[height=3cm]{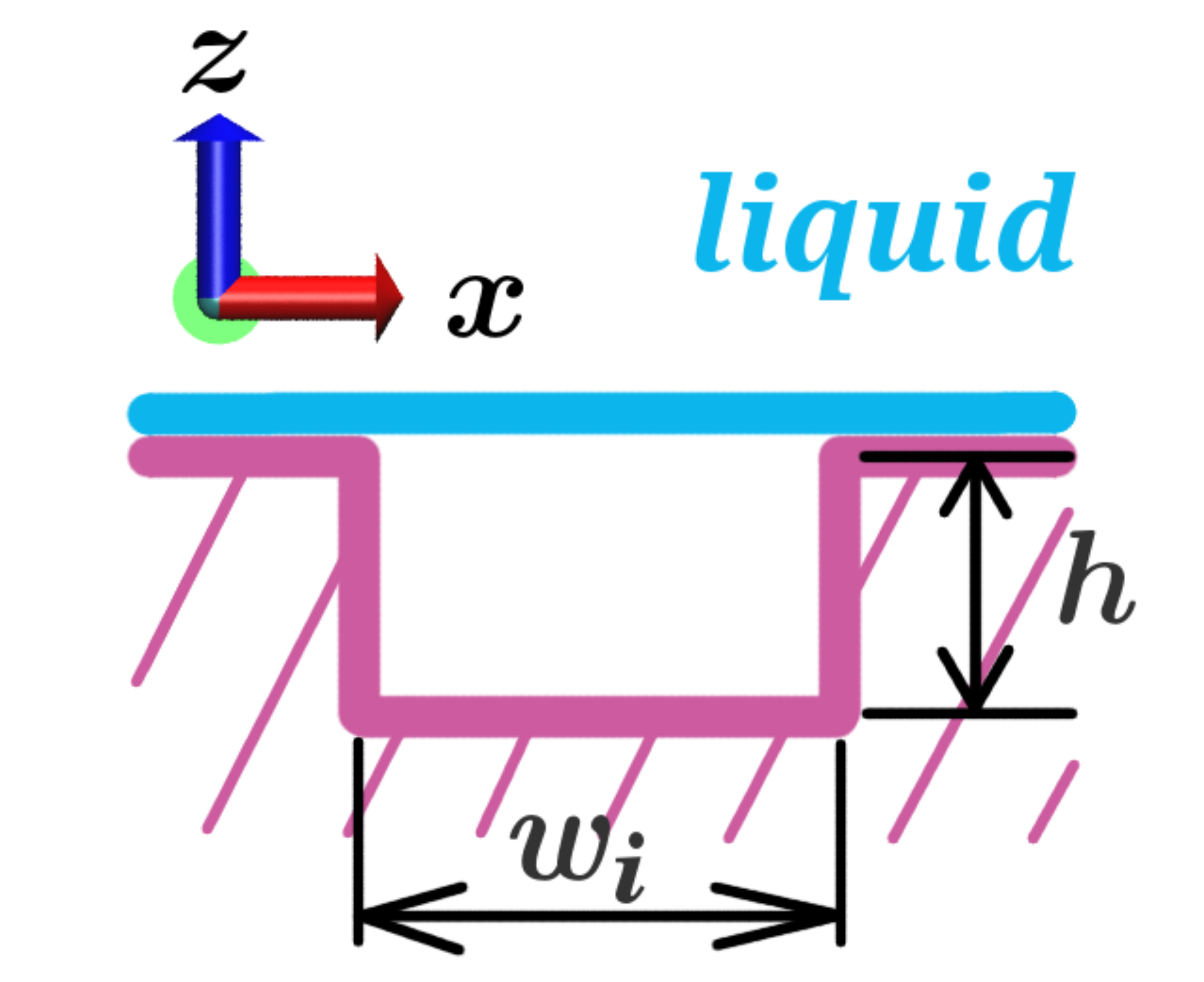}}\hspace{0.2cm}
    \subfloat[]{\label{fig:cav_filled}\includegraphics[height=3cm]{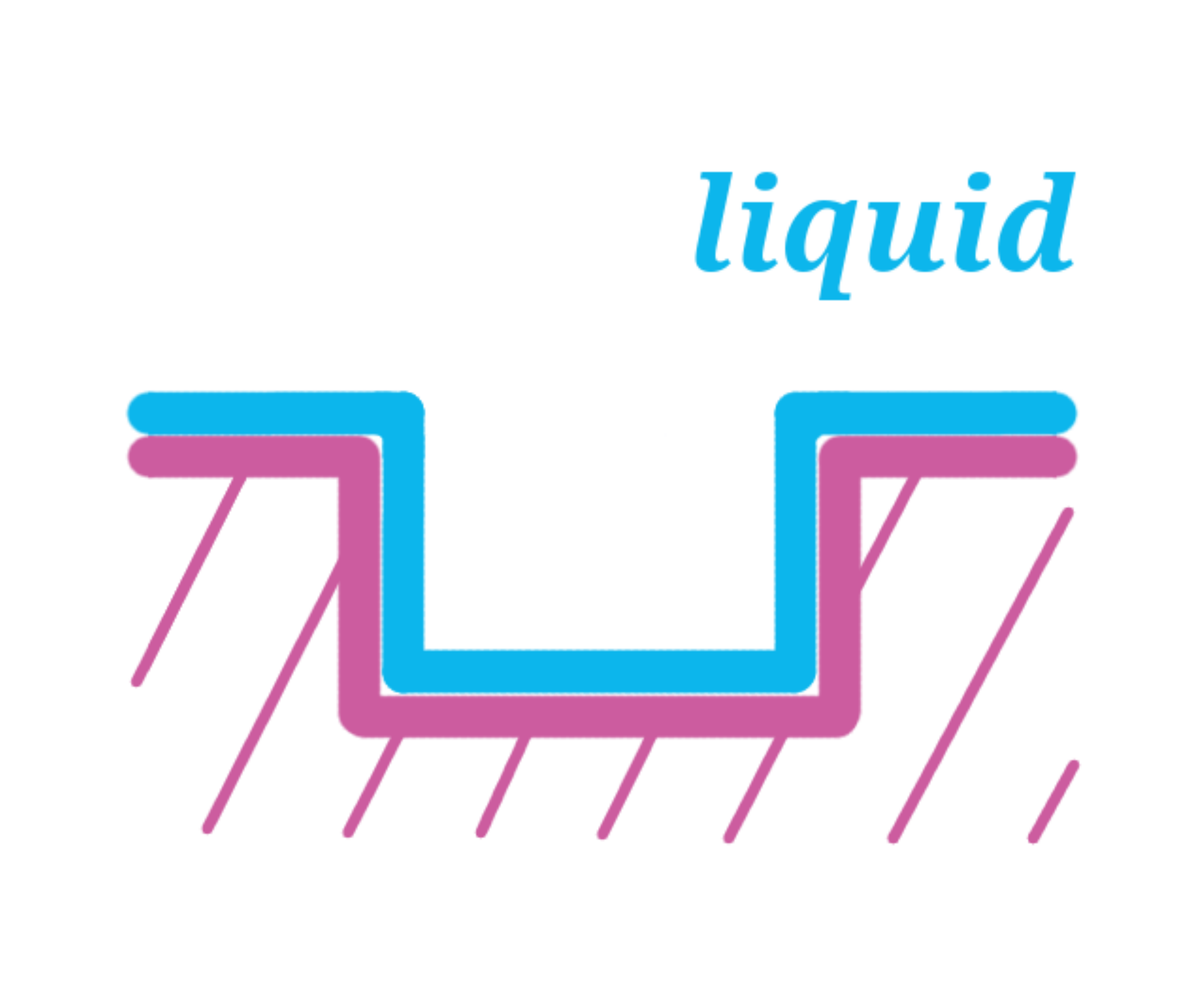}}
    \caption{(Color online) (a) Liquid in Cassie state does not fill the cavity, but creates straight liquid-vapor interface of area {$L_y w_i$}. (b) Liquid in Wenzel state fills the cavity creating solid-liquid interfaces at the bottom and walls of the cavity.}
    \label{fig:cav_unfill_fill}
\end{figure}

The equations of motions are integrated by a velocity-Verlet algorithm using a time step of $\Delta t= 0.005$ in units of $\tau=\sigma\sqrt{m/\epsilon}$, where $m$ denotes the bead mass. A Dissipative Particle Dynamics (DPD) thermostat \cite{PH_JK_1992, PE_PW_1995} with friction coefficient $\gamma_{\rm DPD}= 0.5$ is used to control the temperature at $k_\textrm{B}T = 1.2\epsilon$. The virtue of the DPD thermostat is the local conservation of momentum, which results in hydrodynamic behavior on large time and length scales. The self-diffusion coefficient $D$ of the polymer liquid at coexistence pressure is $D=0.0157\pm0.0030\,\sigma^2/\tau$~\cite{JS_MM_2008}. Thus, the Rouse relaxation time of a polymer is $\tau_\textrm{R}=R_\textrm{ee}^2/(3\pi^2 D)$ is $\tau_\textrm{R} = 25.3 \pm 5\,\tau$.

The flat solid substrate is represented by three layers of face-centered-orthorhombic unit cells of a LJ crystal with number density {of $\rho_\textrm{s} = 2.67 \sigma ^{-3}$}. {The lengths of lattice vectors are $a_x = a_y = \sqrt{3}a$ and $a_z = a$, where $a = \sqrt[3]{0.5}\sigma$}. The atoms of the substrate also interact with liquid particles by a truncated LJ potential $U^\textrm{s}(r)=U^\textrm{s}_\textrm{{LJ}}(r)-U_\textrm{LJ}^\textrm{s}(r_c)$ with 
\begin{equation}
	U_\textrm{LJ}^\textrm{s}(r)=4 \epsilon_\textrm{s} \bigg[ \Big( \frac{\sigma_\textrm{s}}{r}
        \Big)^{12}-\Big( \frac{\sigma_\textrm{s}}{r} \Big)^6 \bigg],
\label{eq:lj_s}
\end{equation} 
where the length scale is set by $\sigma_\textrm{s}=0.75 \sigma$ and the ratio $\epsilon_\textrm{s}/\epsilon$ is changed from $0.2$ to $0.6$ to control the wettability of the substrate. Due to orientation effects the contact angles on the vertical side walls of the grooves slightly differs from the horizontal substrate planes.

The topographically structured substrate has grooves as depicted on {a side view} in Fig.~\ref{fig:struct_sub}. The ratio of the substrate covered by grooves to the projected flat substrate is $\varphi = 0.625$. We study two structured surfaces with the same area fraction $\varphi$. To keep $\varphi$ constant, we vary the distance between the grooves $d$ and the width of the grooves $w$. {These widths of the grooves are taken to be $w_1 = 2.06 \sigma$ and $w_2 = 6.18 \sigma$.} The height of the groove is identical for both substrates $h = 2.38 \sigma$. 

There are two microscopic length scales of the polymer liquid to which the dimensions of the substrate topography can be compared: (i) the end-to-end distance $R_\textrm{ee}=3.43 \sigma$ of the spatially extended chain molecules and (ii) the effective bead size $\sigma$.  The former distance is related to the single-chain conformations in a cavity (cf. Fig.~\ref{fig:cav_unfill_fill}), while the latter sets the intrinsic scale of packing and layering effects in the fluid. For the short chain length considered in the present work, these two scales are not well separated.

The dimensions of the simulation box are $L_x = L_y = 33 \sigma$, whereas $L_z$ is varied from $29.25 \sigma$ to $36.05 \sigma$ depending of the topography of the substrates and strength of solid-liquid interaction $\epsilon_\textrm{s}$. If not mentioned otherwise, $N_0=1920$ chains are confined in the simulation domain. The number of chains is varied in Sec.~\ref{sec:varpress}, where we study flows at different pressures. After equilibration of the initial conformation, we let the system run for $4-8 \times 10^6$ steps to collect statistics. Accounting for the integration timestep $\Delta t$, the length of the simulations is $2-4 \times 10^4 \tau$ or about $10^3$ Rouse relaxation times of the polymers. 

\subsection{Wetting on the topographically structured substrate}
\label{IIb}
Structuring a hydrophobic substrate (contact angle of a drop greater than $90$\textdegree) by grooves, one increases the contact angle and may render the substrate superhydrophobic. When the liquid does not fill the cavities (Cassie state), the contact angle changes according to Cassie's formula \cite{AC_SB_1944}

\begin{equation}
 \cos \theta_\textrm{C} = \varphi \cos \theta_\textrm{E} + \varphi - 1,
 \label{eq:cassie}
\end{equation}
where $\theta_\textrm{C}$ is the contact angle in the Cassie state, and $\theta_\textrm{E}$ is the contact angle on a flat substrate.

In the Wenzel state, the liquid wets the substrate completely and the contact angle is described by Wenzel's formula \cite{Wenzel_1936}
\begin{equation}
 \cos \theta_\textrm{W} = r \cos \theta_\textrm{E},
 \label{eq:wenzel}
\end{equation}
where $\theta_\textrm{W}$ is the contact angle in the Wenzel state, and $r$ stands for roughness of the substrate. The parameter $r$ is the ratio of the actual substrate area wetted by the liquid to the projected area of the substrate. $r \geq 1$ and $r=1$ for a flat unstructured substrate. Our model of the substrate corresponds to the roughness $r=\frac{d+w+2h}{d+w}$, where $w$ is the width of the groove. For the finely and coarsely corrugated substrates the roughness is given by $r_1=1.865$ and $r_2=1.289$, respectively. The values of the contact angles for a flat surface (from anisotropy of the pressure tensor and Young's law~\cite{NT_MM_DT_UT_12}) and the predictions of Cassie and Wenzel formulae (with roughnesses $r_1$ and $r_2$) are compiled in Table~\ref{tab:cont_angles}.

\begin{table}[b]
  \begin{tabular}{l|c|c|c|c}
   $\epsilon_\textrm{s}$ & $\theta_\textrm{E}$, [\textdegree] & $\theta_\textrm{C}$, [\textdegree] & $\theta_\textrm{W}\arrowvert_{r=r_1}$, [\textdegree] & $\theta_\textrm{W}\arrowvert_{r=r_2}$, [\textdegree] \\
   \hline
   $0.2\epsilon$ & $173.5$ & $174.9$ & -- & -- \\
   $0.3\epsilon$ & $154.2$ & $159.7$ & -- & -- \\
   $0.4\epsilon$ & $138.1$ & $147.2$ & -- & $163.5$ \\
   $0.5\epsilon$ & $120.8$ & $134.0$ & $162.8$ & $131.3$ \\
   $0.6\epsilon$ & $102.1$ & $120.4$ & $113.0$ & $105.7$ \\
  \end{tabular}
   \caption{Contact angles,  $\theta_\textrm{E}$, $\theta_\textrm{C}$, and $\theta_\textrm{W}$, on flat  and structured substrates when the liquid is in the Cassie or the Wenzel  states, respectively. Data are obtained from the measured values of the surface and interface tensions of the planar solid-liquid and liquid-vapor tension using Eqs.~(\ref{eq:cassie}) and (\ref{eq:wenzel}) for the two different corrugations $r_1$ and $r_2$.}
   \label{tab:cont_angles}
\end{table}

We emphasize that Cassie and Wenzel predictions, given by Eqs.~(\ref{eq:cassie}) and~(\ref{eq:wenzel}), refer to macroscopic amounts of liquid and do not account for thermal fluctuations of the liquid-vapor interface spanning the grooves or the influence of the three-phase contact line at the edges of the grooves. Since the pinning of a liquid on a patterned substrate occurs in MD simulations, there is a hysteresis of the values of the contact angles. Thus, if one simulates a droplet, the estimates of the contact angle will not necessarily correspond to the Cassie or Wenzel values.

To predict the stability region of the Wenzel state, we use a simple phenomenological continuum model \cite{CCB_JLB_LB_EC_2003}, accounting for free energies of the interfaces presented in Fig.~\ref{fig:cav_unfill_fill}. The liquid will fill the cavity if free energy of the liquid in the Wenzel state (Fig.~\ref{fig:cav_filled}) is smaller than free energy of the liquid in the Cassie state (Fig.~\ref{fig:cav_unfilled}):
\begin{equation}
 F_\textrm{W}-F_\textrm{C} < 0.
\end{equation}

From the geometries of interfaces the free energies can be written as
\begin{equation}
  \begin{array}{c c}
    F_\textrm{C}=\gamma L_y w_i,\\
    F_\textrm{W}=\gamma_\textrm{SL} L_y w_i + \gamma_\textrm{SL} L_y 2 h = \gamma_\textrm{SL} L_y (w_i + 2 h),
  \end{array}
\end{equation}
where $w_i$ is the width of the cavity and $\gamma_\textrm{SL}$ is the solid-liquid surface tension that, in turn, is related to the contact angle of the liquid on a flat substrate $\theta_\textrm{E}$ through Young's equation $\gamma_\textrm{SL}=-\gamma \cos\theta_\textrm{E}$. Here we recall that the density of the vapor is negligible. Therefore we assume a substrate-vapor surface tension of $\gamma_\textrm{SV} \approx 0$. Then, the condition of filling the cavity by the liquid can be formulated in terms of the contact angle $\theta_\textrm{E}$ as
\begin{equation}
 \theta_\textrm{E} < \arccos (-\frac{1}{1+2h/w_i}).
 \label{eq:theta_cont_prediction}
\end{equation}

Substituting $w_i$ with parameters of the substrates of our simulation model, we find that for a finely corrugated substrate the condition for filling the cavity corresponds to $\theta_{\textrm{E}_1} < 108 \text{\textdegree}$, and for a roughly corrugated one it holds $\theta_{\textrm{E}_2} < 124 \text{\textdegree}$. Comparing these phenomenological predictions with values of contact angles from Table.~\ref{tab:cont_angles}, we observe that the filling of cavities of finely and roughly corrugated substrates occurs at $\epsilon_\textrm{s} \approx 0.6 \epsilon$ and $\epsilon_\textrm{s} \approx 0.5 \epsilon$, respectively.

\begin{figure}[ht]
    \centering
    \subfloat[]{\label{fig:popul_s2d1}\includegraphics[height=4.5cm]{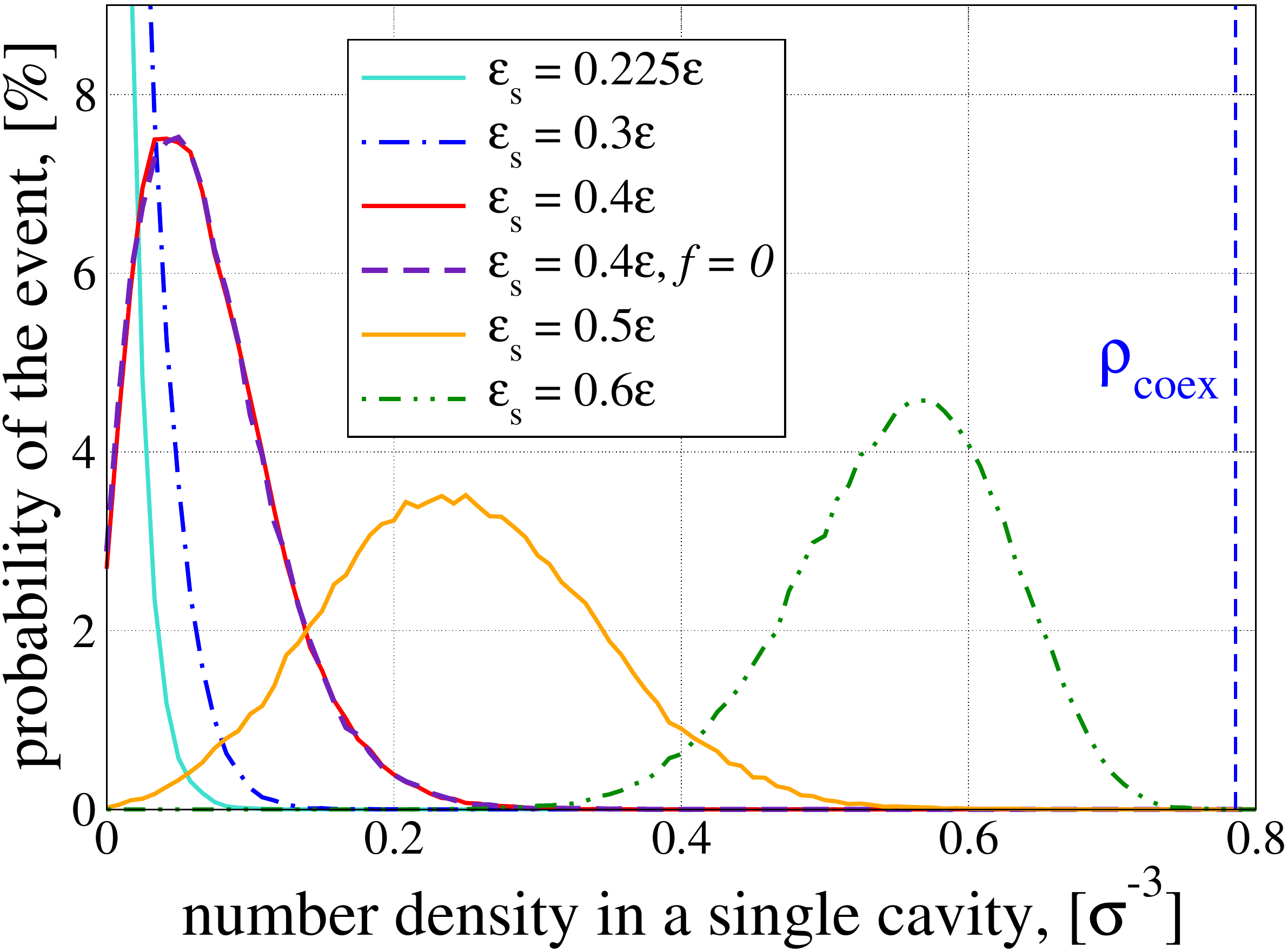}}\vspace*{0.3cm}
    
    \subfloat[]{\label{fig:popul_s7d4}\includegraphics[height=4.5cm]{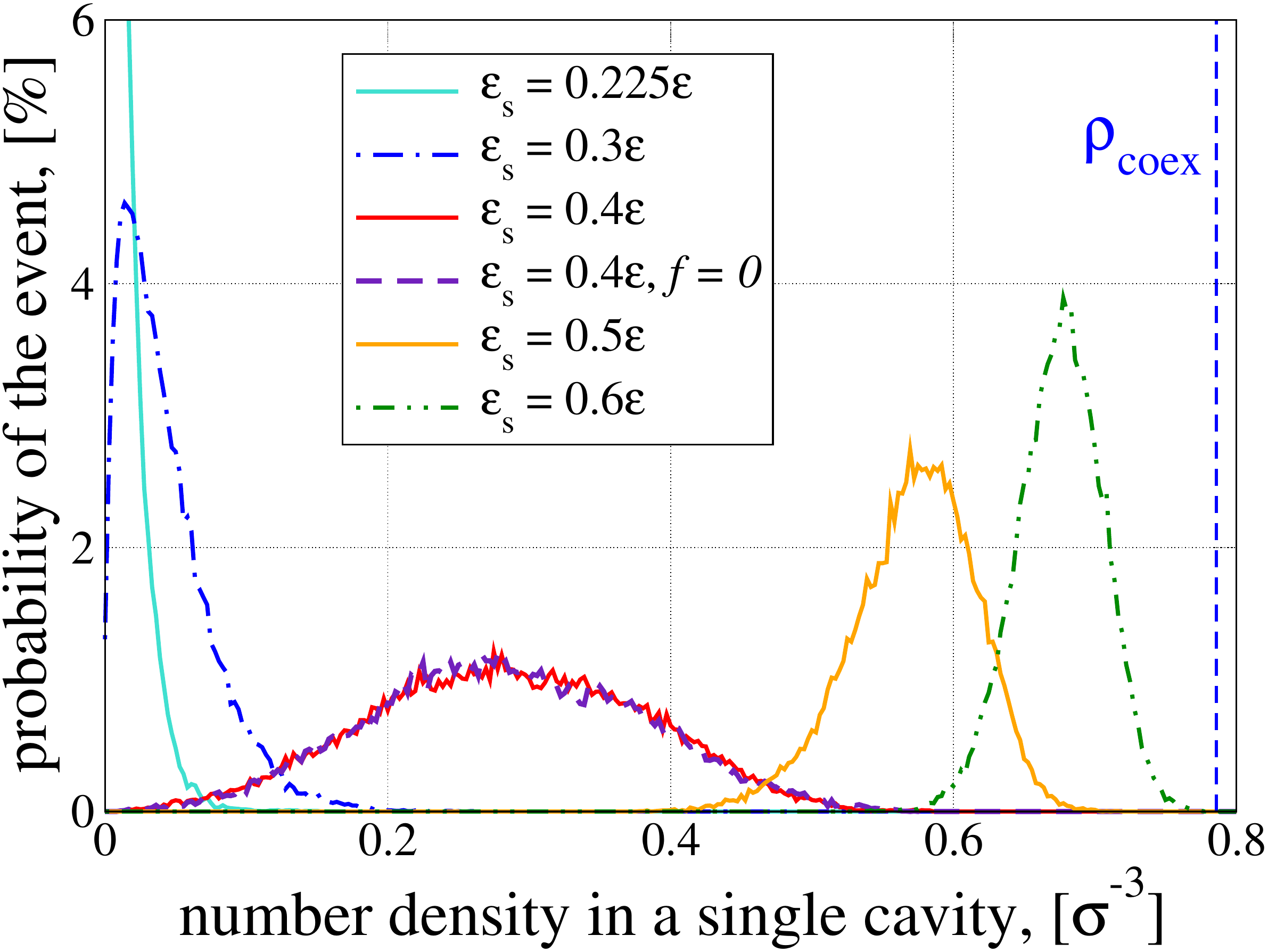}}
    \caption{(Color online) Histogram of a number density in a single cavity for finely (a) and roughly (b) corrugated substrates at varying strengths of solid-liquid interaction $\epsilon_\textrm{s}$ with body force $0.001 \sigma / \tau^2$. Additionally plotted a histogram in the crossover regime at $\epsilon_\textrm{s}=0.4\epsilon$ without body force (dashed lines). The presence of a body force exerts only negligible influence on the density distribution. The coexistence density $\rho_\textrm{coex}$ is plotted by dashed vertical line.}
    \label{fig:population}
\end{figure}

This phenomenological, macroscopic consideration predicts a first-order surface phase transition between the Cassie and the Wenzel state. In Fig.~\ref{fig:population} we present the number density of liquid in a single cavity of the substrate for the two topographies. If there were a phase transition of first order between the Cassie and Wenzel states, one would expect a bimodal histogram, where the two peaks corresponded to empty cavities without liquid (Cassie state) and cavities completely filled by the liquid (Wenzel state), respectively. In marked contrast, however, we observe only a single peak in the density distribution. As we vary the solid-liquid interactions, $\epsilon_{\textrm{s}}$, the average number density of liquid beads inside the cavity changes gradually. Around half-filling ($\rho_\textrm{coex}/2 \approx 0.4\sigma^{-3}$) the distributions are very broad. The coexistence density of the liquid is indicated in the graph by the dashed vertical lines. 
The value inside the filled cavity is comparable but it remains smaller due to packing (layering) of the particle fluid inside the cavity. Snapshots reveal that the liquid-vapor interface strongly fluctuates in this crossover region and adopts states between straightly spanning the top of the grooves, as expected for the Cassie state, {and} touching the bottom of the groove, as being characteristic for the Wenzel state. In passing, we note that the application of the body force to the liquid does not change the behavior {(cf.~solid and dashed lines in Fig.~\ref{fig:population} in presence and absence of a body force at $\epsilon_{\textrm{s}}=0.4\epsilon$, respectively)}.   

There are two effects that contribute to the absence of a true surface phase transition in our simulations: (i) If the coupling between neighboring cavities were negligible because the cavities were widely separated, the number of particles in each groove would be an independent quantity. Then, since each groove would be a quasi-one-dimensional system, there cannot be a true thermodynamic phase transition. Instead, we would expect to observe large but finite-sized domains of filled and unfilled portions along the groove. This rational is important for widely spaced and extremely long grooves, and this mechanism will eliminate all thermodynamic singularities at finite temperatures. 
While a systematic study of the filling/emptying behavior as a function of the spatial extension along the groove is beyond the scope of the present study, it is unlikely that this effect causes the absence of a sharp thermodynamic transition for the rather short grooves in our simulations because we do not observe multiple domains along a groove.

(ii) The absence of bimodal probability distributions in Fig.~\ref{fig:population} indicates that for studied substrates the transition between the Wenzel and the Cassie state is not of first-order but continuous.
Thermal fluctuations give rise to fluctuations of the particle number in a cavity due to the finite compressibility of the liquid and, more importantly, due to fluctuations of local position of the liquid-vapor interface inside the groove. The latter are the analog of capillary waves in this confined geometry~\cite{Andy,AP_CR_NW_2007,RR_AP_2011}. For the substrates used in our simulation study, the height of the grooves is of the same order of magnitude as the intrinsic width of the liquid-vapor interface, and already small excursions of the interface position result in a significant change of the filling fraction. 
Even if $d/\sigma$ is not very small but the grooves are widely separated to be considered independent, we anticipate large fluctuations because the macroscopic consideration, Eq.~(\ref{eq:theta_cont_prediction}), asserts that the Wenzel and Cassie state have equal free energies for $\theta_{\textrm{E}} \to 90$\textdegree. In this case, the liquid-vapor interface can move homogeneously like a rigid plane up and down in the groove. 

However, we point out that if the grooves become deeper, one should expect the first-order transition from the Cassie to the Wenzel state, in accord with macroscopic considerations.

If the liquid is confined into a channel, the pressure might deviate from the vapor pressure, at which the liquid and vapor of vanishingly small density coexist. The associated effects will be discussed in the following section.

%
\subsection{Normal pressure}
\label{sec:norm-press}
%

We study a polymer liquid that is confined between two apposing substrates. A flat unstructured substrate, a finely corrugated substrate with roughness $r_{1}$ and a substrate with a rough corrugation, $r_{2}$, are considered. Additionally, we vary the strength of solid-liquid interaction, $\epsilon_\textrm{s}$. To adjust the normal pressure to its coexistence value, $p_\textrm{coex} \approx 0$, we change the distance $L_z$ between the two apposing substrates and calculate the corresponding pressure inside the liquid composed of $N_0=1920$ chains. The resulting width, $L_z^\textrm{coex}$, is therefore a function of solid-liquid interaction strength $\epsilon_\textrm{s}$, the corrugation $r_i$, and the number of chains $N_0$, i.e., $L_z^\textrm{coex} = L_z^\textrm{coex}\,(\epsilon_\textrm{s},\,r_i,\,N_0)$.

We calculate the normal pressure tensor component $p_\textrm{n}$~\cite{AllenTild89, FrenkSmit02} dividing the systems into slabs perpendicular to the direction of the substrates and using the approach by Irving and Kirkwood \cite{JI_JK_1950}. The pressure is divided into a configurational term, which is proportional to the number density $\rho(k)$ in a slab $k$, and a virial term:
\begin{equation} 
p_\textrm{n}(k)=k_\textrm{B}T\langle\rho(k)\rangle+\frac{1}{V_\textrm{sl}}\Big\langle{\sum_{i<j}}^{(k)}\frac{z_{ij}^2}{r_{ij}}|\mathbf{F}_{ij}|\cdot\cos{(\widehat{\mathbf{F}_{ij},\mathbf{r}_{ij}})}\Big\rangle,
\end{equation}
where $V_\textrm{sl}$ denotes the volume of the slab, $x_{ij}, y_{ij}, z_{ij}$ are the distances between interacting particles $i$ and $j$ in $x, y, z$ directions, respectively, and $r_{ij}^2 = x_{ij}^2 + y_{ij}^2 + z_{ij}^2$.  The angle between the force $\mathbf{F}_{ij}$ and the distance vector $\mathbf{r}_{ij}$ is given by expression $\widehat{\mathbf{F}_{ij},\mathbf{r}_{ij}}$, and its cosine can take the values $\pm 1$. Angular brackets $\langle \cdots \rangle$ denote averages in the canonical ensemble. The sum $\sum_{i<j}^{(k)}$ runs over particles $i$ and $j$ if a portion of line connecting them is located inside the slab $k$. The interactions between the solid substrate and the liquid particles also contribute to the virial \cite{NT_MM_DT_UT_12}. 
Since the solid-liquid interaction is of finite range, however, they do not contribute to the pressure in the center of the confined film. Mechanical stability asserts that the pressure normal to the surfaces does not depend on position and thus equals the pressure at the center of the film.

\begin{figure}[ht]
 \centering\vspace*{0.2cm}
 \subfloat[]{\label{fig:press_flat}\includegraphics[height=5cm]{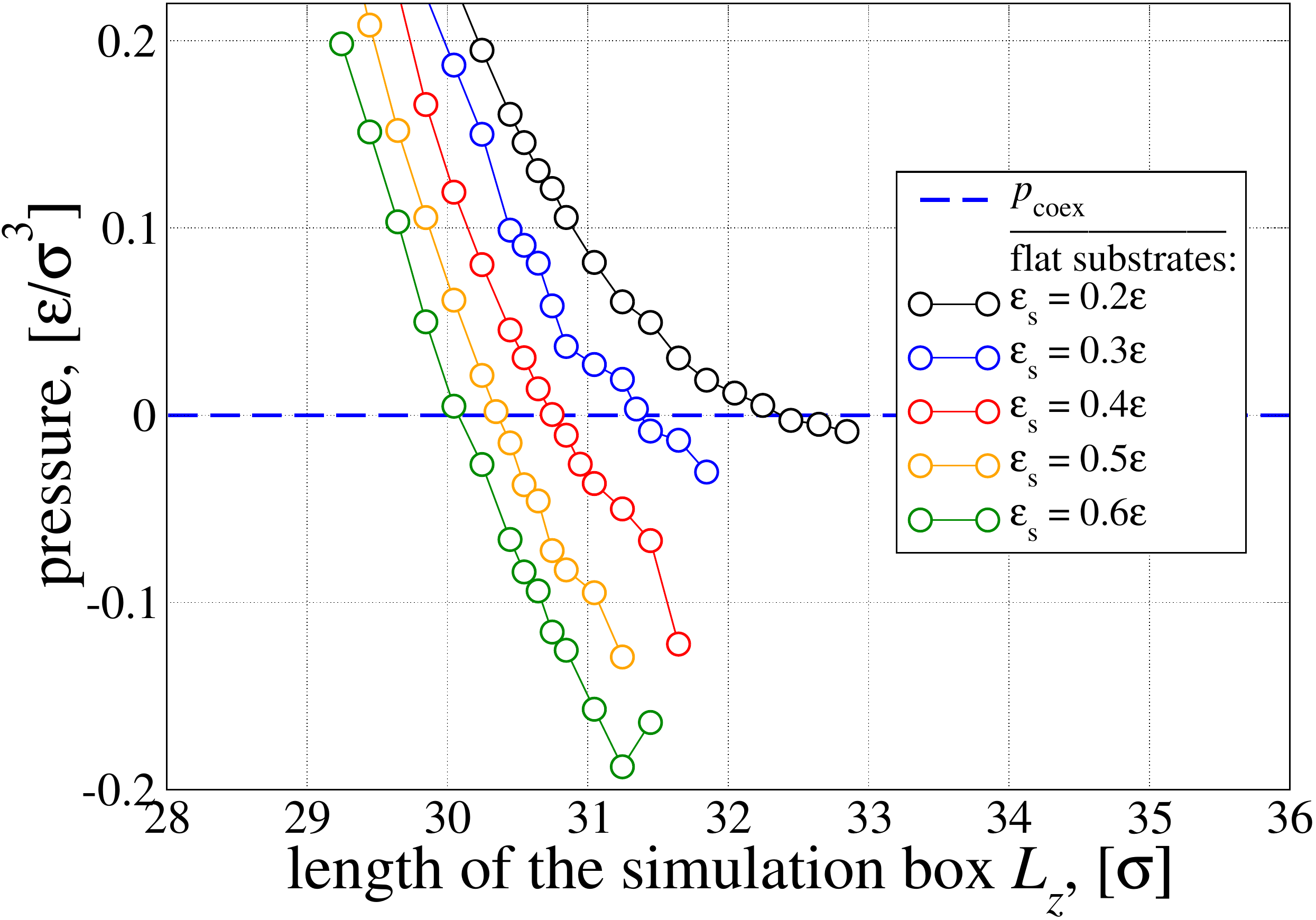}}\vspace*{0.5cm}
 
 \subfloat[]{\label{fig:press_s2d1_s7d4}\includegraphics[height=5cm]{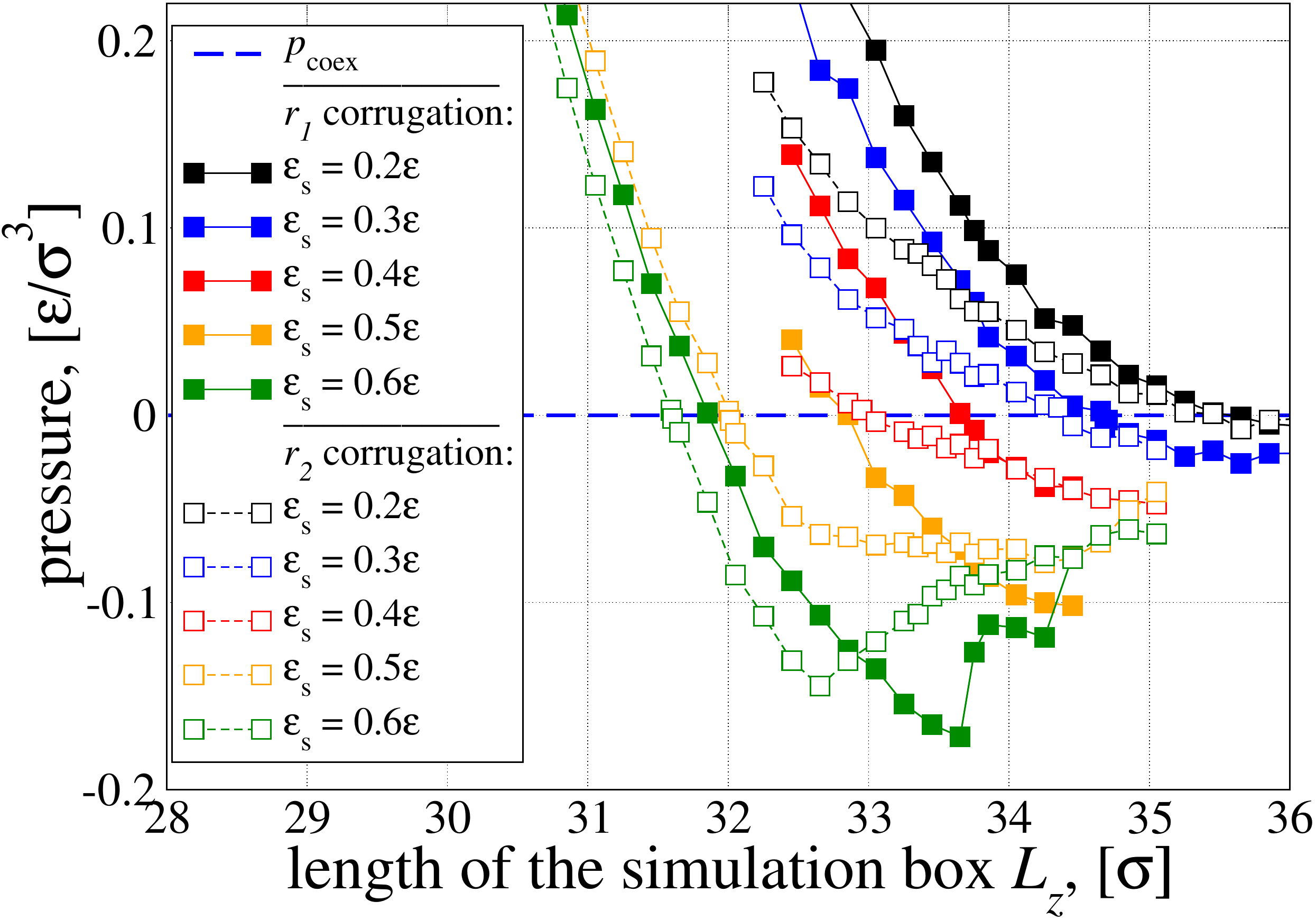}}
 \caption{(Color online) Pressure of the polymer liquid at the center of the film. Panel (a) shows the results for a flat surface, while panel (b) presents the results for the two corrugated substrates. Filled symbols represent the corrugation $r_1$ and empty symbols correspond to $r_2$. Dashed horizontal lines indicate the coexistence density and pressure, respectively.}
 \label{fig:comp_dens_press}
\end{figure}

The dependence of the normal pressure on the distance, $L_{z}$, between the walls is presented in Fig.~\ref{fig:comp_dens_press}. At small $L_{z}$, the density and pressure at the center of the film monotonously decrease with increasing $L_{z}$. At constant number of particles, we observe that the higher the attractive strength of solid-liquid interaction, $\epsilon_\textrm{s}$, is the smaller is $L_{z}^\textrm{coex}$, because of the excess of the compressible liquid at the attractive substrate. $L_{z}$ is defined as the maximal distance between the corrugated substrates, i.e., measured from the bottoms of the corrugation. $L_{z}^\textrm{coex}$ is larger for a finely corrugated substrate with roughness $r_{1}$ than for the coarse grooves (filled {\it vs.} open symbols in Fig.~\ref{fig:comp_dens_press}), because the polymers explore the wider grooves better, and thereby they increase the liquid excess at the substrate and reduce the pressure at the center.

At large wall separation,  $L_z > 32.5\sigma$, we find deviations from the expected monotonous dependence on $L_{z}$ due to the formation of vapor bubbles under tensile stress, $p<0$. These cavitation events nucleate more readily at {strongly} attractive surfaces $\epsilon_\textrm{s} \geqslant 0.5 \epsilon$ and for corrugated substrates. In those cases, the simulations did not necessarily achieve equilibrium.

%
\section{Hydrodynamic boundary condition}
\label{sec:flat-vs-pattern}
%
%
\subsection{Navier's slip condition and computational techniques}
\label{sec:Navier}
%

In order to study flow past these substrates, we move the two apposing substrates into opposite directions with a small absolute velocity $v_\textrm{wall}^\textrm{C}= 0.075 \sigma/\tau$ (Couette flow). The average shear rate is defined as a ratio of the velocity of the wall $v_\textrm{wall}^\textrm{C}$ to the width of the channel $L_z$ and, in our simulations, it is of the order of $\dot{\gamma} \sim 5\cdot10^{-3}\,\tau^{-1}$. The product of the shear rate and the Rouse relaxation time, $\tau_{\textrm{R}}$, of the polymers defines the Weissenberg number, $\textrm{Wi} = \dot{\gamma}\tau_\textrm{R}$. Since $\textrm{Wi} \approx 0.13 \ll 1$, we do not expect the weak flow to significantly perturb the molecular conformations and, indeed, density profiles obtained with and without flow are almost identical (see e.g.~Fig.~\ref{fig:population}).

When the fluid flows past the substrate, the velocity profile deviates from the macroscopic behavior, and the velocity of the liquid at the surface may not equal that of the substrate, i.e., the liquid slips past the surface. These surface-induced deviations from the macroscopic behavior can be incorporated into a macroscopic continuum description via boundary conditions. A popular choice is Navier's hydrodynamic boundary condition \cite{Navier_1823}

\begin{equation}
 \eta \frac{\partial v_x}{\partial z} \bigg{|}_{z=z_\textrm{h}} = \lambda v_\textrm{s},
 \label{eq:navier}
\end{equation}
where $\eta$ denotes the viscosity of the liquid, $v_x$ is the $x$ component of the velocity of the liquid parallel to the substrate. $\lambda$ is the friction coefficient, and the slip velocity, $v_{s}=v_x(z_\textrm{h})$, represents the velocity of the liquid at the surface. If applied to the detailed microscopic velocity profile, this equation would quantify the balance between the viscous stress in the liquid and the friction stress of the liquid slipping past the surface. In order to serve as a boundary condition of a continuum description (i.e., the Navier-Stokes equation), however, changes of the liquid structure at the surface are ignored, and $\eta$ is interpreted as the shear viscosity of the liquid in the bulk, $\eta=5.3 \pm 0.1 \sqrt{m \epsilon} / \sigma^2$~\cite{JS_MM_2008}. 
By the same token, $v_{x}(z)$ is interpreted as the macroscopic velocity profile that obeys the continuum description and that we can extract from simulations by extrapolating the linear profile of Couette flow towards the substrate, $z=z_\textrm{h}$.

Thus, Navier's hydrodynamic boundary condition, Eq.~(\ref{eq:navier}), parameterizes the flow past the surface by two effective material constants of the surface~\cite{JLB_LB_1999,JLB_LB_1999_FD,MM_CP_2008,JS_MM_2008,JS_MA_FS_2008}: the hydrodynamic position, $z_\textrm{h}$, at which the boundary condition to the macroscopic continuum description is applied and the slip length, $\delta=\eta/\lambda$. The flow of simple liquids over flat substrates is often described by the no-slip boundary condition, $\delta=0$ \cite{LB_JLB_2007}, and the hydrodynamic position, $z_\textrm{h}$, often coincides with the location of the sharp solid-liquid interface, as intuitively expected. Finite slip has been observed in complex liquids \cite{MM_CP_2008, MM_CP_JS_2008, JS_MM_2008, OB_KJ_2010, FL_JS_CP_MM_2011} or at specific substrates: superhydrophobic \cite{CCB_SJ_JB_2002, CCB_JLB_LB_EC_2003, SG_YZ_HL_2003, CCB_CB_EC_2004, CL_CHC_CLK_2008, SS_AB_JH_OV_2012} and chemically patterned ones~\cite{TQ_XPW_PS_2005, Priezjev_2011}.

\begin{figure}[t]
    \centering
    \subfloat[]{\label{fig:cou_s7d4_e05}\includegraphics[height=3.5cm]{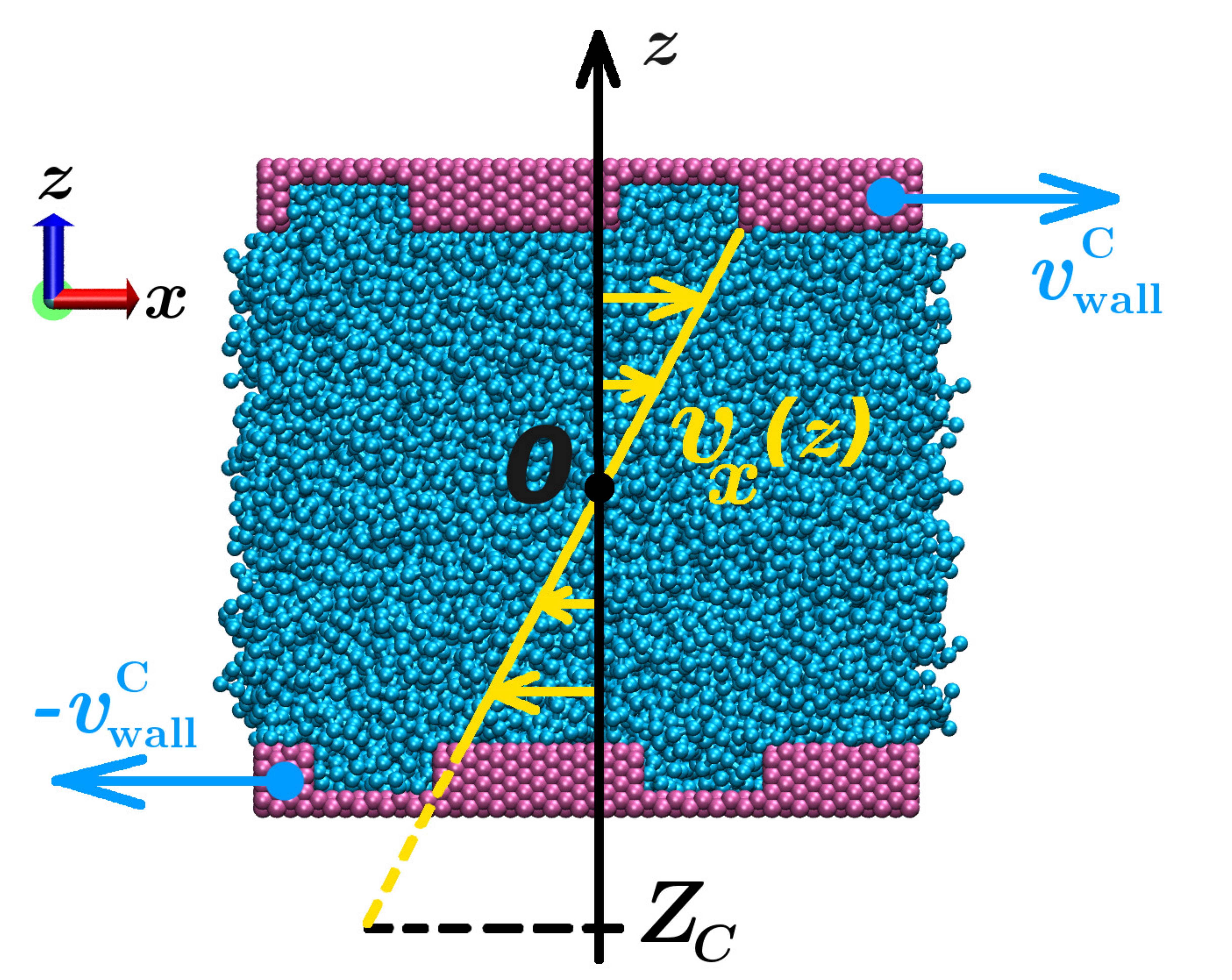}}\hspace*{0.1cm}
    \subfloat[]{\label{fig:poi_s7d4_e05_f5}\includegraphics[height=3.5cm]{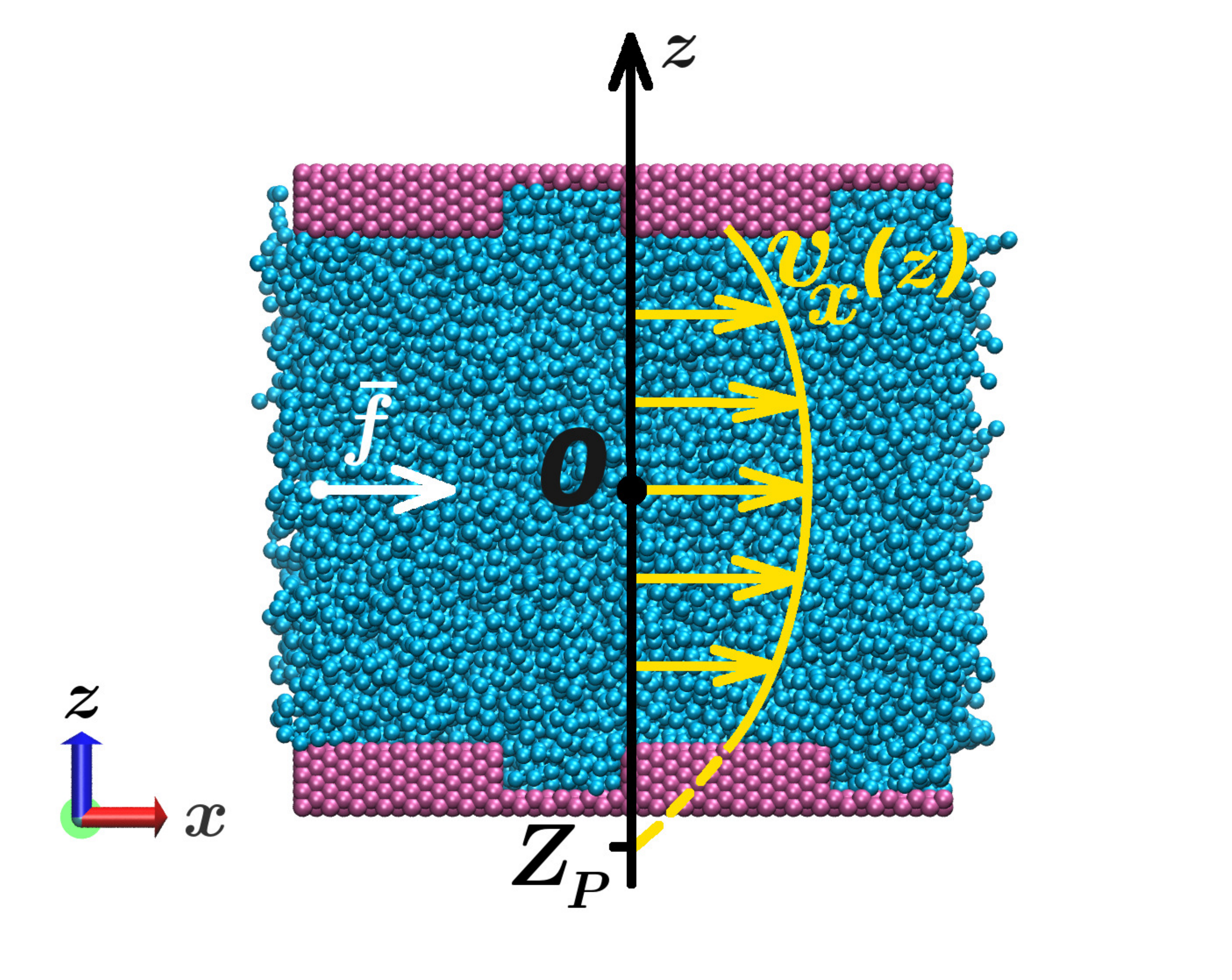}}
    \caption{(Color online) (a) Couette flow created in the system with roughly corrugated substrates. Two walls are moved into opposite directions with absolute velocities $v_\textrm{wall}^\textrm{C}$. The origin of the $z$-axis is taken in the middle of the channel (black). The linear velocity profile is sketched by the arrows of different size. The interpolated profile reaches the velocity of the wall at the coordinate $Z_\textrm{C}$. (b) The flow created by application of a body force $\vec{f}$ (white) onto the particles of the liquid. The parabolic velocity profile (arrows) is interpolated to the velocity of the wall ($v_\textrm{wall}^\textrm{P}=0\:\sigma/\tau$) and reaches it at $Z_\textrm{P}$.}
    \label{fig:flows}
\end{figure}

Here we study the flow of a polymer liquid past corrugated substrates. Therefore we expect a finite slip length, $\delta$, and the effective location, $z_\textrm{h}$, at which the hydrodynamic boundary conditions is to be applied, is not obvious. Since Navier's boundary condition features two independent parameters, we use two flow profiles: Couette flow, generated by moving the surfaces, and Poiseuille-like flow, generated by applying an external body force on the liquid. The linear and parabolic velocity profiles expected from the macroscopic continuum model for Couette and Poiseuille flow, respectively, are illustrated in Fig.~\ref{fig:flows}.

We chose the origin of the coordinate system at the center of the film. The macroscopic velocity profile is linear $v_x^\textrm{C}(z)=az+b$ for Couette flow, and the constants $a$ and $b$ are determined from the boundary conditions $v_x^\textrm{C}(0)=0$ and $v_x^\textrm{C}(Z_\textrm{C})=-v_\textrm{wall}^\textrm{C}$, where $Z_\textrm{C} < 0$ is the position where extrapolation of the linear, macroscopic velocity profile equals to the velocity of the bottom wall $-v_\textrm{wall}^\textrm{C}$ as shown in Fig.~\ref{fig:cou_s7d4_e05}. The resulting velocity profile is therefore
\begin{equation}
 v_x^\textrm{C}(z)=-\frac{v_\textrm{wall}^\textrm{C}}{Z_\textrm{C}}z.
 \label{eq:cou}
\end{equation}

According to the macroscopic Navier-Stokes equation, the velocity profile  of the flow generated by an external body force has a parabolic shape, $v_x^\textrm{P}(z)=-\frac{\rho_\textrm{coex} f}{2 \eta}z^2+Az+B$, where $\rho_\textrm{coex}=0.786 \sigma^{-3}$ is the number density of the liquid and $\eta$ denotes its viscosity. $A=0$ by symmetry. When we express the constant $B$ by the boundary condition $v_x^\textrm{P}(\pm Z_\textrm{P})=0$, we obtain:

\begin{equation}
 v_x^\textrm{P}(z)=\frac{\rho_\textrm{coex} f}{2 \eta}(Z_\textrm{P}^2 - z^2).
 \label{eq:poi}
\end{equation}
$Z_\textrm{P}$ is the coordinate, where the extrapolated parabolic velocity profile of the liquid reaches the velocity of the wall 
as illustrated in Fig.~\ref{fig:poi_s7d4_e05_f5}. Velocity profiles are fitted by Eqs.~\ref{eq:cou} and \ref{eq:poi} at the center of the film excluding the near-substrate region. Applying the hydrodynamic boundary condition, Eq.~(\ref{eq:navier}), to the macroscopic profiles, Eqs.~(\ref{eq:cou}) and (\ref{eq:poi}) we relate the extrapolated positions $Z_\textrm{C}$ and $Z_\textrm{P}$, which we extract from the simulations, to the slip length, $\delta=\sqrt{Z_\textrm{C}^2-Z_\textrm{P}^2}$, and boundary position, $z_\textrm{h} = Z_\textrm{C} + \delta$ \cite{MM_CP_2008,Servantie08b}.

To generate Poiseuille-like flow, we impose body forces $f$ of strength $0.001 \sigma / \tau^2$, $0.003 \sigma / \tau^2$, $0.005 \sigma / \tau^2$ and $0.008 \sigma / \tau^2$. An external force injects energy into the system, which is removed by the thermostat, keeping the temperature constant. In order to preserve a linear response of the liquid, only the smallest force was applied at small values of $\epsilon_\textrm{s}$, whereas for $\epsilon_\textrm{s} \geqslant 0.4 \epsilon$ all four forces were used.
%
\subsection{Position of the hydrodynamic boundary}
\label{sec:slip-frict}
%

\begin{figure}[bt]
    \centering
    \includegraphics[height=4.5cm]{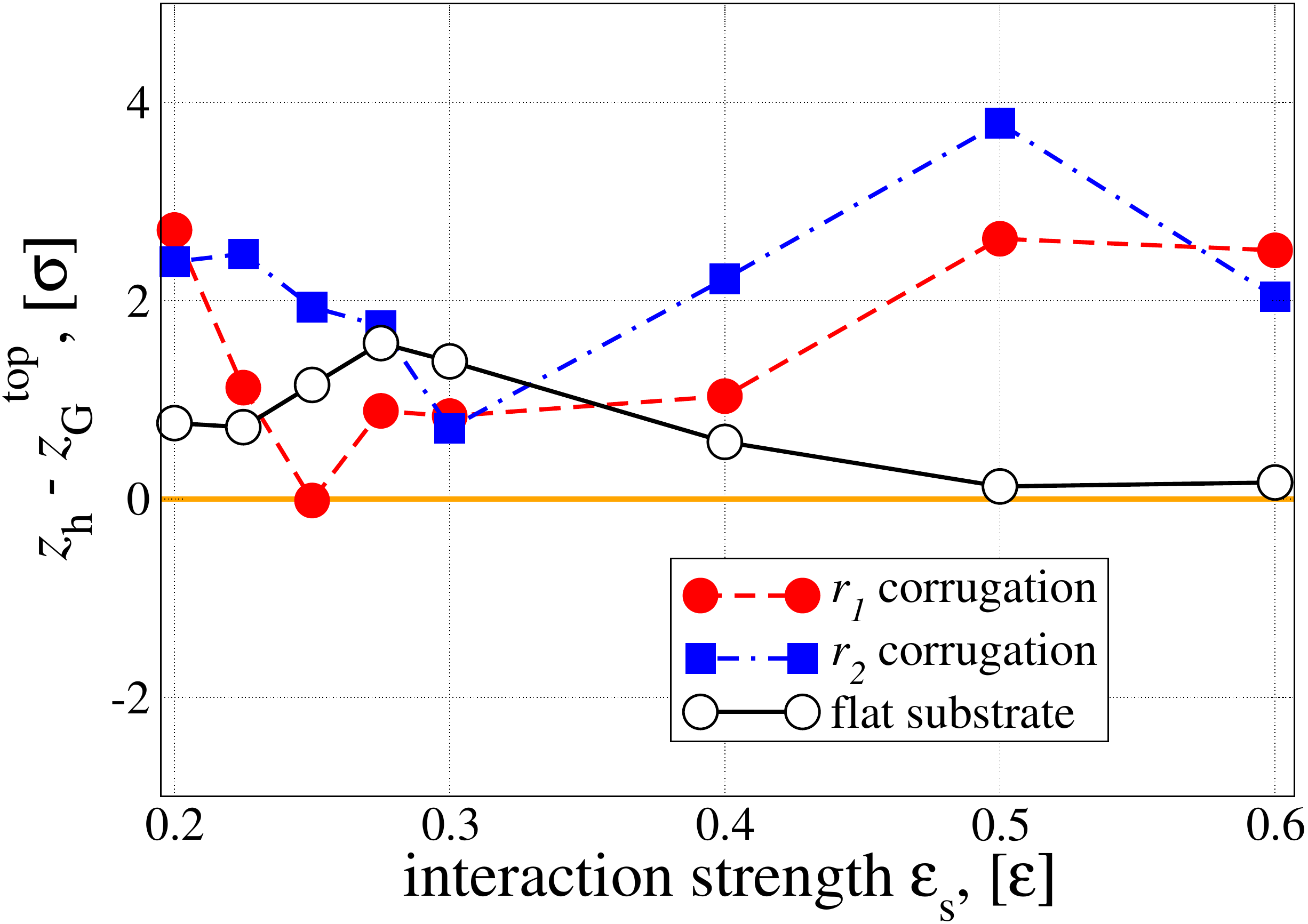}
    \caption{(Color online) Distance of the position, $z_\textrm{h}$, at which Navier's boundary condition~\ref{eq:navier} is to be applied, to the innermost substrate atom $z_\textrm{G}^\textrm{top}$ for channels with fine and rough corrugations (lines with filled circles and squares, respectively) and for channels with flat walls (line with open circles). Positive values, $z_\textrm{h} - z_\textrm{G}^\textrm{top} > 0$, indicate that the effective boundary position is closer to the center of the channel than any particle of the wall.}
    \label{fig:s2d1_s7d4_b_pos_vs_eps}
\end{figure}

Using the simulation data obtained from Couette and Poiseuille-like flows at $p_\textrm{coex} \approx 0$, we can independently determine the slip length, $\delta$, and the position of the hydrodynamic boundary, $z_\textrm{h}$. In Fig.~\ref{fig:s2d1_s7d4_b_pos_vs_eps} we present the distance between the boundary position $z_\textrm{h}$ and the position, $z_\textrm{G}^\textrm{top}$, of the innermost substrate segment, {which is located on the tops of ridges between the grooves. In case of a flat substrate (black line with open circles), the position $z_\textrm{G}^\textrm{top}$ refers to the innermost atomic layer of the substrate. It is interesting to notice, that even in the case of flat walls and small $\epsilon_\textrm{s}$, the hydrodynamic boundary is found somewhat inside the liquid, but not directly at the solid-liquid interface}. 
For most parameters at structured substrates, the hydrodynamic boundary is located $2-4$ segment diameters inside the liquid above the top of the ridges between the grooves. This observation for our molecular fluid even holds in the Wenzel state, where the liquid enters the grooves. 

\begin{figure}[t]
    \centering
    \includegraphics[height=2.5cm]{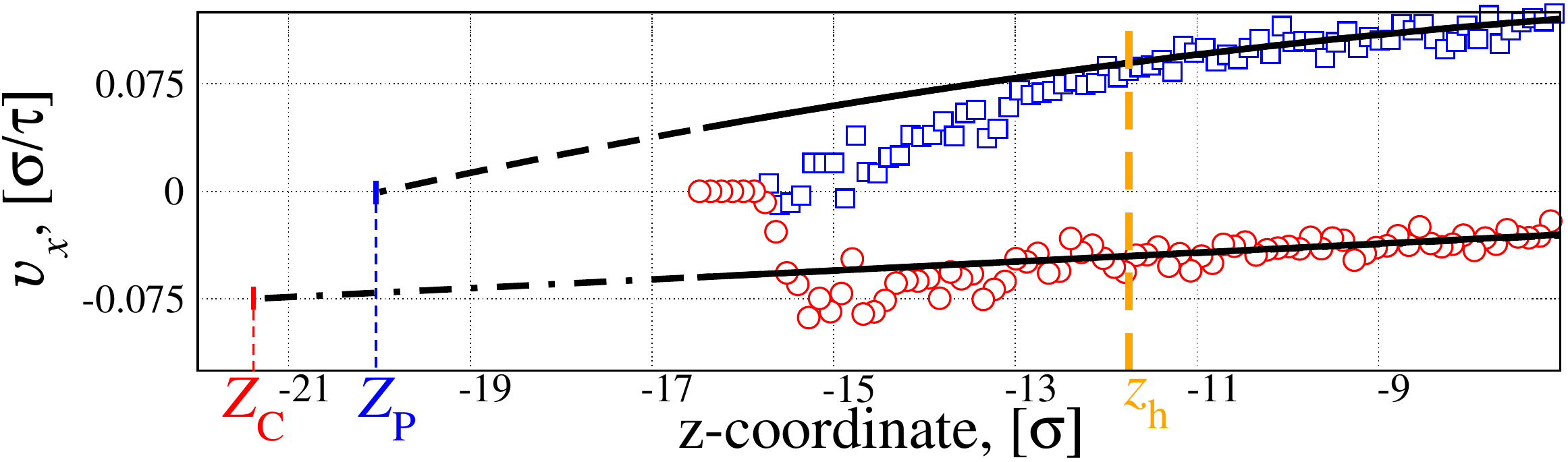}
    \caption{(Color online) Velocity profiles of Couette flow (circles) and Poiseuille-like flow with body force $f=0.005\sigma/\tau^2$ (squares) at $\epsilon_\textrm{s} = 0.4\epsilon$ and rough corrugation $r_2$ are shown. The case corresponds to the crossover regime, when liquid enters the cavity. Smooth thick solid lines are the fits to the velocity profiles, measured away from the substrate. The positions $Z_\textrm{C}$ and $Z_\textrm{P}$, where extrapolated macroscopic profiles reach the velocity of the wall, $v_\textrm{wall}^\textrm{C}= -0.075 \sigma/\tau$ and $v_x^\textrm{P}(-Z_\textrm{P})=0$ for Couette and Poiseuille-like flows, respectively, are indicated. These no-slip positions depend on the type of the flow and therefore cannot be used as a boundary condition.
    }
    \label{fig:z_h_Z_P_Z_C}
\end{figure}

The position of an effective no-slip plane, where the extrapolation of the hydrodynamic velocity field, vanishes -- $Z_\textrm{C}$ for Couette flow and $Z_\textrm{P}$ for Poiseuille flow -- will be located at an intermediate coordinate between the peaks and valleys of the substrate topography in the Wenzel state if the slip length, $\delta$, is larger than $2-4$ segment diameters. This result is compatible with  lattice Boltzmann simulations of a continuum fluid past statistically rough surfaces \cite{CK_JH_OV_2010}. We emphasize, however, that the no-slip position depends on the flow and, hence, cannot be used to parameterize a boundary condition, as shown in Fig.~\ref{fig:z_h_Z_P_Z_C}. Moreover, upon reaching the limit $\theta_\textrm{E}\rightarrow90\text{\textdegree}$ the slippage can be less than $2$ segment diameters (cf. section~\ref{sec:comp-coex}), and the corresponded no-slip position shifts into the liquid.

\begin{figure*}
    \centering
    \subfloat[]{\label{fig:s2d1_e04_flux}\includegraphics[height=5cm]{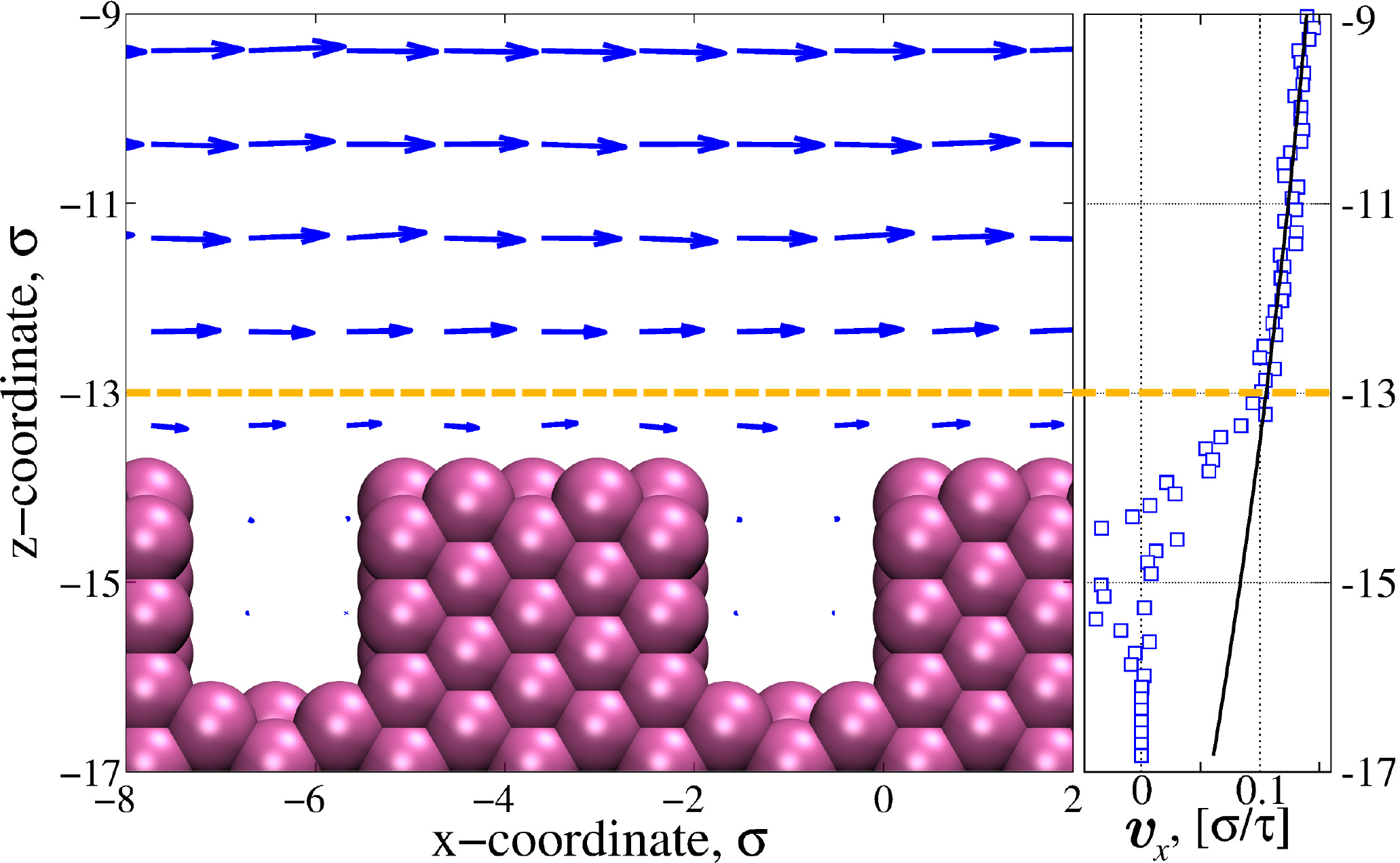}}\hspace*{0.3cm}
    \subfloat[]{\label{fig:s7d4_e04_flux}\includegraphics[height=5cm]{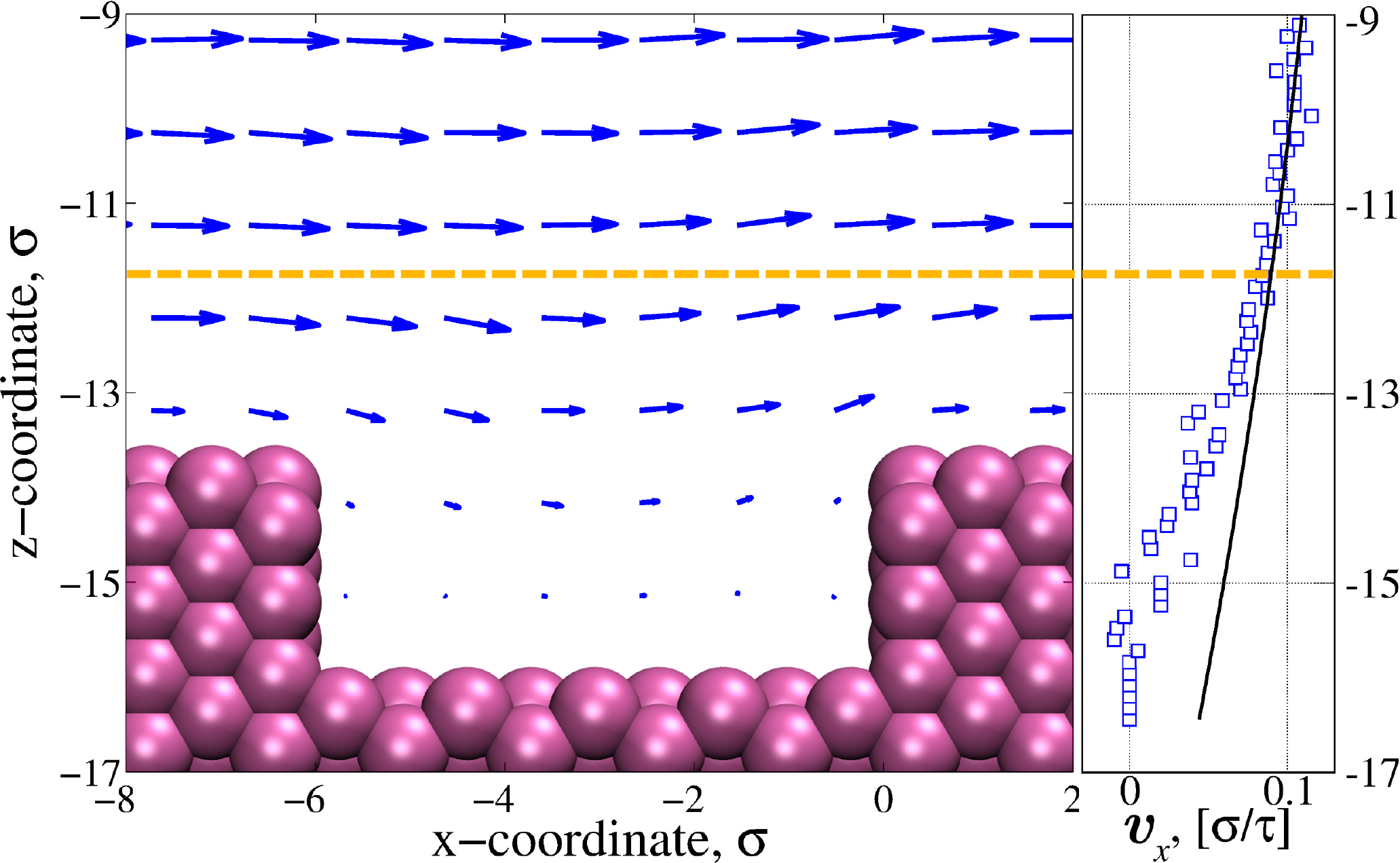}}
    \caption{(Color online) Mass flux, $\rho \vec{v}$ (arrows), in the $xz$ plane at $\epsilon_\textrm{s} = 0.4\epsilon$ and $w_\textrm{1} = 2.06\sigma$ in panel (a)  and at $\epsilon_\textrm{s} = 0.4\epsilon$ and $w_\textrm{2} = 6.18\sigma$ in panel (b) under body force $f=0.005\sigma/\tau^2$ (Poiseuille-like flows). Both cases, (a) and (b), correspond to the crossover regime, when liquid enters the cavity. In the accompanying plots velocity profiles (squares) and the extrapolated hydrodynamic profile (thick solid line) are shown.
    Smooth solid lines are the fits to the velocity profiles, measured away from the substrate. The position of the hydrodynamic boundary, $z_\textrm{h}$, is indicated by horizontal dashed lines.}
    \label{fig:fluxes}
\end{figure*}

In Fig.~\ref{fig:fluxes} we present the mass-weighted, two-dimensional velocity profiles for corrugated substrates at $\epsilon_\textrm{s} = 0.4\epsilon$. Additionally, we demonstrate the velocity profiles $v_x$ (noisy solid lines) and their macroscopic fits (smooth solid lines) measured away from the substrate to neglect distortion by the friction at the boundary.
The position of the hydrodynamic boundary condition is also indicated in the graph (horizontal dashed line). In both cases, it is above the corrugation. Qualitatively, $z_\textrm{h}$ describes the crossover between two behaviors: For larger distances from the substrate, the velocity is strictly parallel to the $x$-axis as predicted by the Navier-Stokes equation for Poiseuille and Couette flow past a macroscopic substrate, while closer to the substrate, the velocity field acquires a spatially periodic, perpendicular $z$-component. In agreement with detailed hydrodynamic calculations of flow past a corrugated substrate, the amplitude of the perpendicular velocity component will decay faster away from the substrate if the lateral period is smaller \cite{OV_AB_2011}. 
Additionally, we note that the deviations of the velocity profiles from the macroscopic fits begin at the vicinity of hydrodynamic boundary position. 

Finally, we mention that, at $\epsilon_\textrm{s}=0.4\epsilon$ (crossover regime), the average velocity of the liquid in a cavity of the finely corrugated substrate nearly vanishes, whereas in case of a rough corrugation the liquid actually flows in the cavity. This effect may influence the local shear stress and the effective boundary properties.

%
\subsection{Slip length and surface friction}
\label{sec:comp-coex}
%

The second parameter of the hydrodynamic boundary condition is the slip length, $\delta$, or, equivalently, the friction coefficient $\lambda=\eta/\delta$. If the macroscopic continuum description of the velocity profile were accurate up to the position of the hydrodynamic boundary, $z_\textrm{h}$, and the liquid properties did not change in the vicinity of the substrate, then we could interpret the friction coefficient as the proportionality coefficient between the microscopic friction stress and the local velocity of the fluid flowing past the substrate. 

\begin{figure}[t]
    \centering
    \includegraphics[height=5cm]{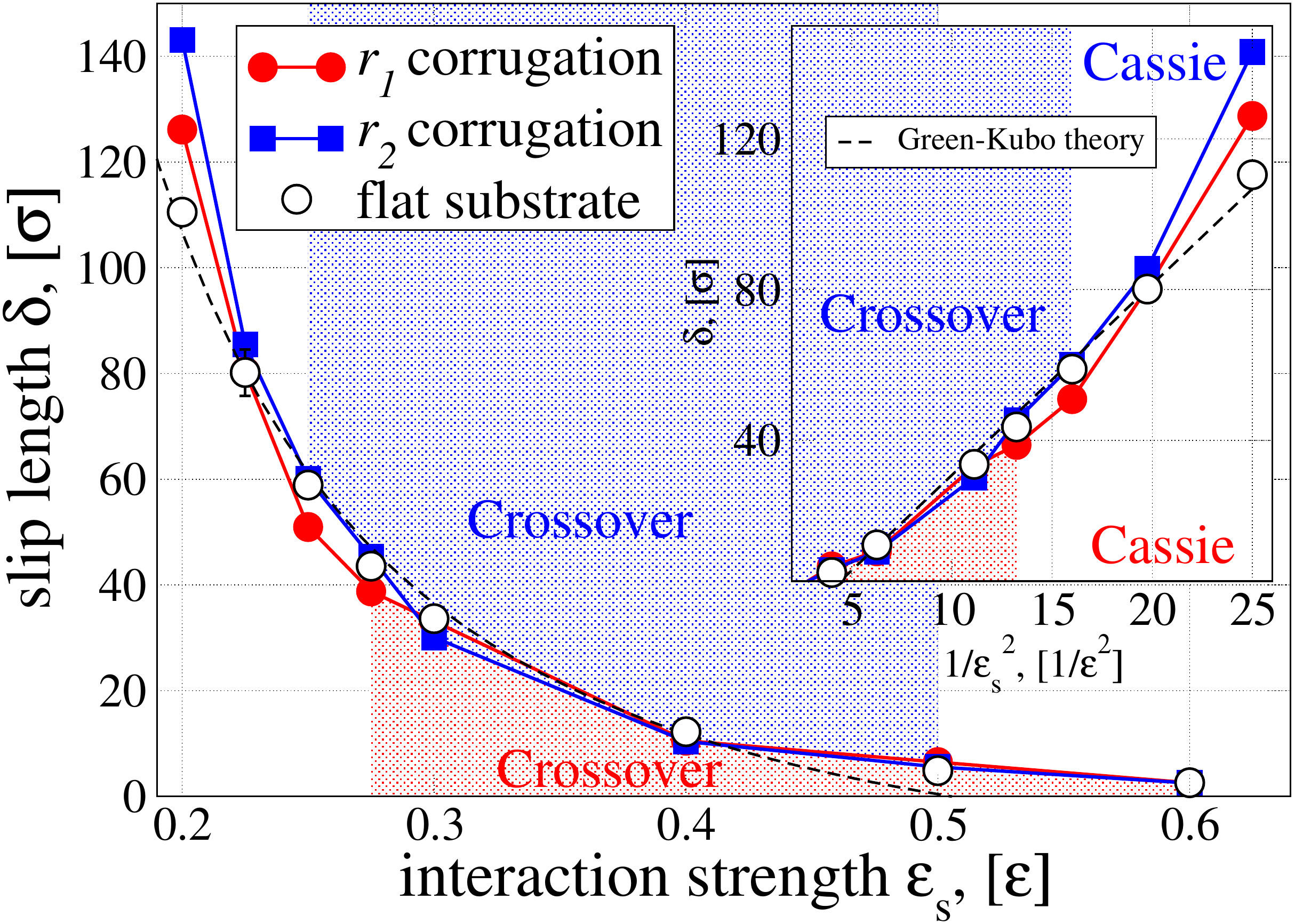}
    \caption{(Color online) The slip length $\delta$ for a liquid at flat (open circles) and patterned substrates of $r_1$ and $r_2$ corrugations (filled circles and squares, correspondingly) at coexistence pressure. The inset shows the slip length $\delta$ as a function of inverse squared interaction strength $\epsilon_\textrm{s}$. }
    \label{fig:slip_comp}
\end{figure}

In this case, one can compute the friction coefficient via a Green-Kubo relation for the autocorrelation function of the tangential force, $F_{x}$, between the liquid and the substrate \cite{JLB_LB_1999_FD}, 

\begin{equation}
\frac{\eta}{\delta} = \lambda= \frac{1}{k_\textrm{B} T A} \int_0^{+\infty}{\rm d}t\; \langle F_x(t) F_x(0)\rangle
\end{equation}
where $A=L_x L_y$ is the area of the substrate. The analysis of Barrat and Bocquet \cite{JLB_LB_1999_FD} suggests that $\lambda$ be proportional to the square of the attraction between liquid and substrate, $\lambda \sim \epsilon_\textrm{s}^{2}$. Previous simulations for a Lennard-Jones polymer liquid on a Lennard-Jones solid corroborated this relation \cite{FL_JS_CP_MM_2011}, and we also observe this dependence on $\epsilon_\textrm{s}$ for our flat substrate as demonstrated in Fig.~\ref{fig:slip_comp}. Since the orientation of the unit cell of the crystalline substrate structure differs from our previous work, the friction coefficient, $\lambda_\textrm{flat}$, that we observe is smaller than for the previous substrate. Such an orientation dependence of friction has also been observed in other studies \cite{CS_TY_PT_2007}.

In this subsection we aim to adjust the pressure to its coexistence value, $p_{\rm coex}\approx 0$, such that the liquid-vapor interface in the Cassie state horizontally spans the grooves. Then, one expects that the friction in the Cassie state is reduced, $\lambda = \varphi \lambda_\textrm{flat}$, compared to its value on a flat substrate, where $\varphi$ is the fraction of the substrate area in contact with the liquid \cite{CCB_JLB_LB_EC_2003}.

In the Wenzel state, the macroscopic expectation is that the friction, like the surface free energy, is proportional to the microscopic contact area, $r$, between liquid and substrate. Thus a finely corrugated substrate, $r_{1}=1.865$ is expected to give rise to more friction (with respect to the projected substrate area) than a coarsely corrugated one, $r_{2}=1.289$.

Fig.~\ref{fig:slip_comp} presents the slip length, $\delta$, as a function of the solid-liquid attraction. The qualitative behavior is similar for flat and corrugated substrates. For strong attraction, the slip lengths is microscopic but upon approaching $\epsilon_\textrm{s} \to 0$ (purely repulsive substrate), the contact angle approaches $180$\textdegree (drying transition) and a lubricating thin vapor layer intervenes between the substrate and liquid giving rise to large slippage. 

\begin{figure}[t]
    \centering  
    \includegraphics[height=5cm]{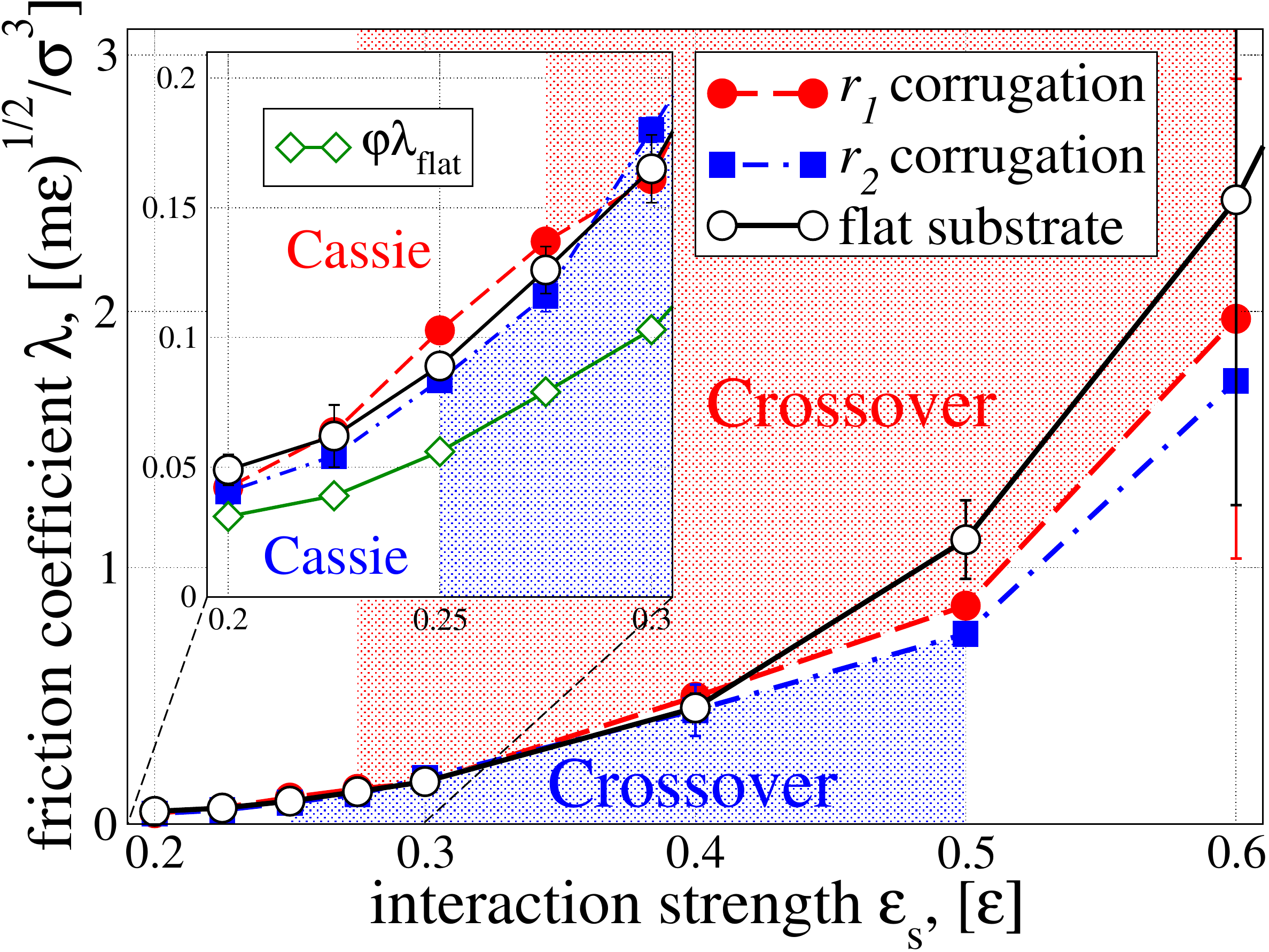}
    \caption{(Color online) Friction coefficient $\lambda$ for different topographies of the substrate. The line with diamonds in the inset corresponds to the friction at the flat substrate multiplied by the area fraction $\varphi=0.625$ covered by grooves. It is smaller than measured friction coefficients for corrugated substrates due to the friction at the edges of the grooves. The crossover regions, separating Cassie and Wenzel states, are shown by shaded areas for the different corrugations.
    }
    \label{fig:frict_comp}
\end{figure}

When plotting $\lambda \sim 1/\delta$ {\it vs.} $\epsilon_\textrm{s}^{2}$, one expects to find that the simulation data are compatible with two linear relations with different slopes. One linear relation corresponds to the Wenzel state for large $\epsilon_\textrm{s}$, and another linear relation with a larger slope characterizes the Cassie regime, where the substrate has less contact with the liquid. In our simulation, however, the crossover gradually occurs in the wide range of $\epsilon_\textrm{s}/\epsilon \in (0.275; 0.6)$ or $\epsilon_\textrm{s}/\epsilon \in (0.25; 0.5)$ for the finely and coarsely corrugated substrate. Only at $\epsilon_\textrm{s} = 0.6 \epsilon$ or $\epsilon_\textrm{s} \geqslant 0.5 \epsilon$, respectively, we observe the system consistently in the Wenzel state. Therefore, we cannot identify the linear behavior that marks the Wenzel state.

In the crossover region, the friction coefficients of both substrates do not significantly differ as shown in Fig.~\ref{fig:frict_comp}. According to macroscopic considerations, we expected the finely corrugated substrate to generate more friction because of the larger contact area with the fluid. In our simulations, however, the liquid is rather trapped inside the finely corrugated grooves, and the friction is not predominately generated at the solid-liquid interface but by viscous dissipation due to the velocity gradient at the interface between the trapped liquid inside of the groove and the flowing liquid at the center. 

In the Cassie state, the friction of the corrugated substrate is comparable to that of the flat substrate. The macroscopic prediction that the friction is reduced by a factor of $\varphi=0.625$ is not observed in our model. Moreover, the friction on the finely corrugated substrate is slightly higher than on the coarsely corrugated one. This effect can be partially rationalized by the effect of edges.

\begin{figure}[t]
    \centering
    \subfloat[]{\label{fig:pack_s2d1_e03}\includegraphics[height=2.44cm]{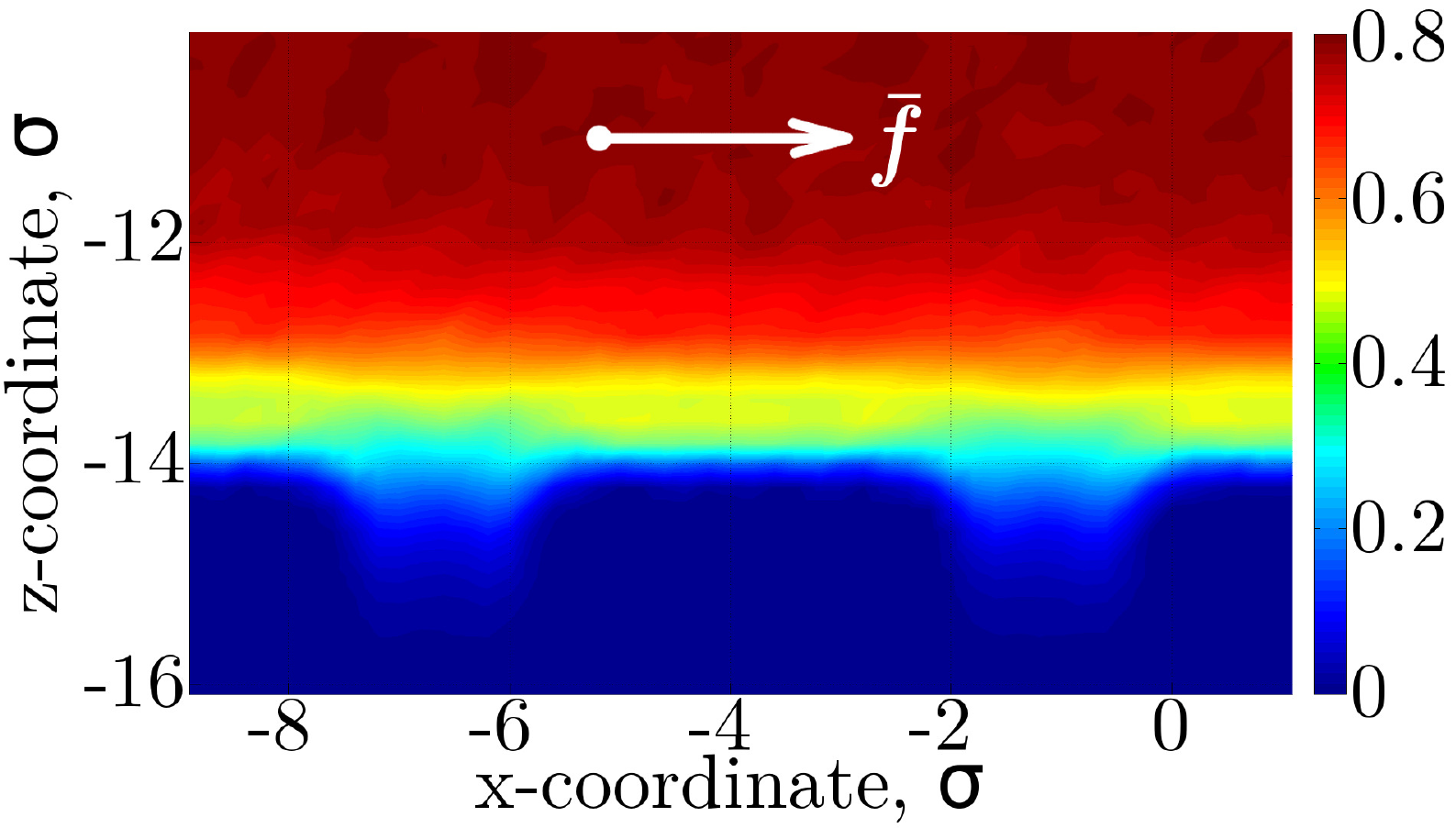}}\hspace*{0.03cm}
    \subfloat[]{\label{fig:pack_s7d4_e03}\includegraphics[height=2.44cm]{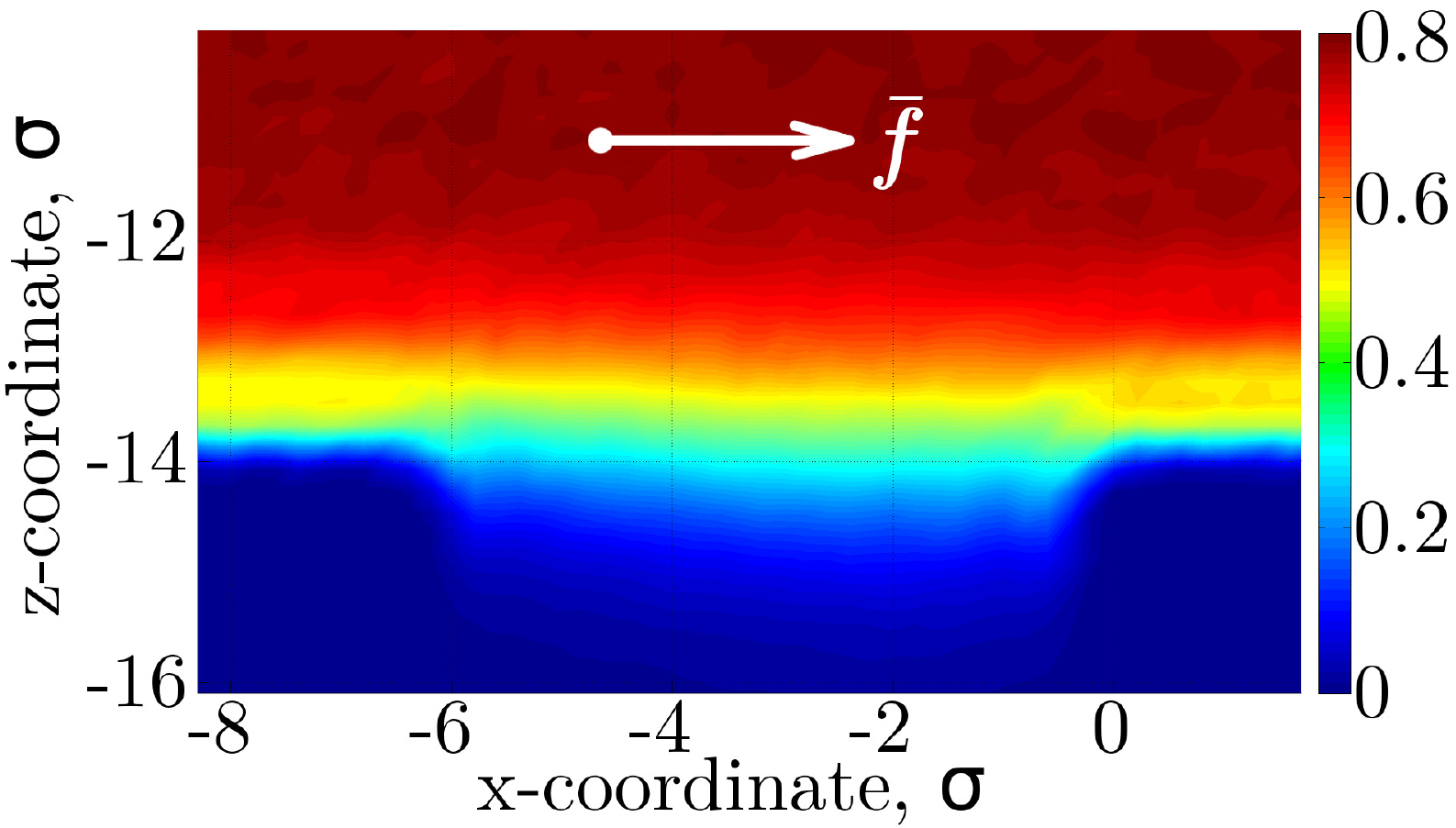}}\vspace*{0.0cm}
    
    \subfloat[]{\label{fig:pack_s2d1_e04}\includegraphics[height=2.44cm]{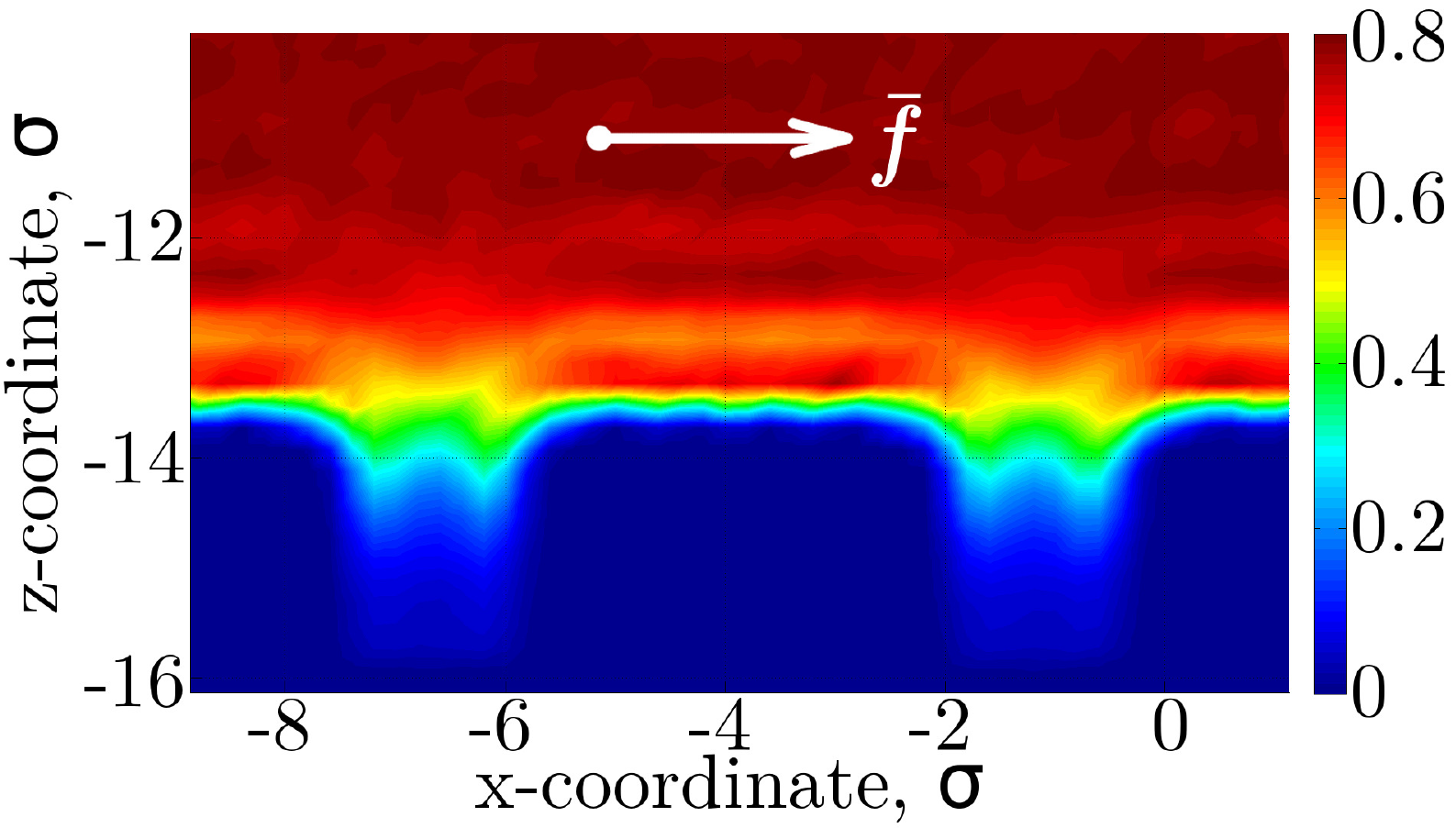}}\hspace*{0.03cm}
    \subfloat[]{\label{fig:pack_s7d4_e04}\includegraphics[height=2.44cm]{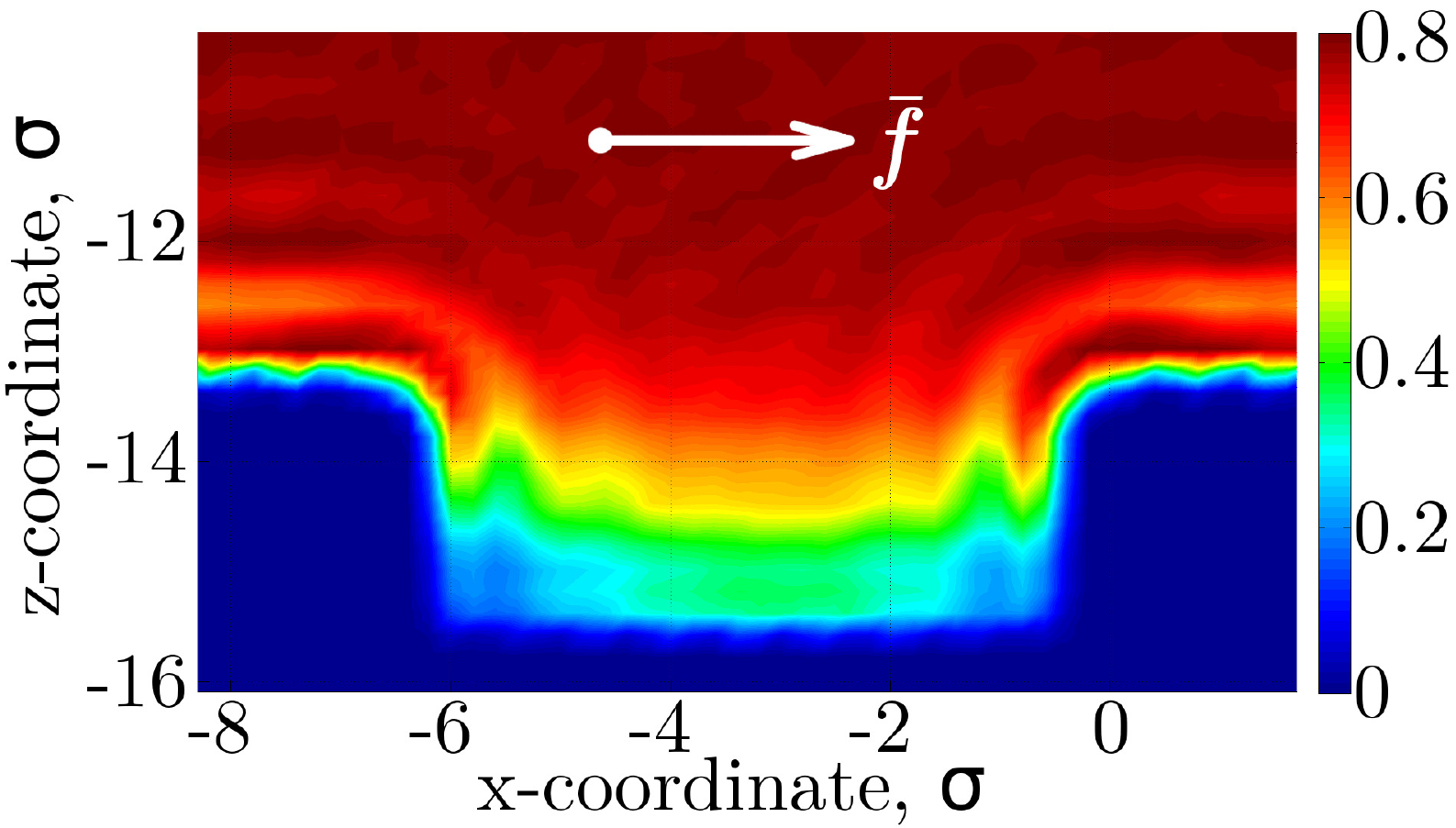}}
    \caption{(Color online) Two-dimensional map of the number density of the particle liquid in the vicinity of the finely (left column) and coarsely (right column) corrugated substrate for $\epsilon_\textrm{s}=0.3\epsilon$ (top row) and $\epsilon_\textrm{s}=0.4\epsilon$ (bottom row). A body force $f=0.005\sigma/\tau^2$ generates a Poiseuille-like flow whose direction is indicated by the arrow.} 
    \label{fig:packing}
\end{figure}

In Fig.~\ref{fig:packing} we present two-dimensional density plots in the vicinity of the corrugated substrate. We observe that density oscillations that indicate layering effects of the particle fluid emerge from the edges of the grooves and propagate a few particle diameters, $\sigma$, into the channel. We also note that even for the rather small body force, $f=0.005\sigma/\tau^2$, that drives the fluid through the channel, the flow breaks the left-right symmetry, and the packing effects at the left and right edges of a groove differ slightly. The edge facing the flow (on the right side of the cavity) exhibits stronger layering effects than the edge on the left side. 

To a first approximation, we try to quantify these deviations from the macroscopic prediction due to localized perturbations of the flow by augmenting the Cassie model of friction by a term that accounts for additional friction at the edges of the grooves. 

\begin{equation}
 \lambda = \varphi \lambda_\textrm{flat} + K(\epsilon_\textrm{s},d,w) \Sigma_\textrm{edge},
 \label{eq:fric_pinned}
\end{equation}
where $\Sigma_\textrm{edge}=\frac{2}{d+w}$ is the line density of edges and $K(\epsilon_\textrm{s},d,w)$ quantifies the additional friction coefficient per edge length. $K$ slightly depends on the geometry but in the limit of isolated edges, $d,w \to \infty$, we expect $K$ to adopt a finite value. The line density of edges is three-fold higher for the finely corrugated substrate than for the coarsely corrugated one. From the data of Fig.~\ref{fig:frict_comp} we extract the effective edge friction $K$.  The values are in the range of $0.03 \leqslant K \leqslant 0.16$ for the finely corrugated substrate and somewhat larger, $0.09 \leqslant K \leqslant 0.22$, for the wider grooves. The order of magnitude of the edge friction, $K \sim 2.5 \lambda_{\rm flat} \sigma$, is plausible, i.e., each edge generates an additional friction that corresponds to roughly two additional rows of the substrate. 
The increase of $K$ with increasing the distance between the edges is in accord with Fig.~\ref{fig:packing}, where we observe that liquid layering effects are more pronounced for wider grooves than for more narrow ones. 

%
%
\subsection{Pressure-driven flow in patterned channels}
\label{sec:varpress}
%
%

For an incompressible liquid confined into a channel with flat boundaries, the flow that is generated by a body force and pressure-driven Poiseuille flow are equivalent. For corrugated substrates, however, both methods of generating flow are no longer equivalent: If we use a body force to set the fluid in motion, the system will remain translationally invariant along the flow direction, and the pressure at the center of the channel will be independent from the position along the channel. If we generate the flow by a pressure gradient, instead, the pressure will decrease along the channel. 
Even if the liquid were incompressible and hence its shear viscosity would not depend on pressure, we expect that the parameters of the hydrodynamic boundary condition of a corrugated substrate -- slip length, $\delta$, and position, $z_\textrm{h}$ -- will depend on the pressure and therefore will vary along the channel. The ability of the fluid to explore the cavities of the substrate makes the total system effectively compressible. Upstream, where the pressure is high, the fluid is more likely to enter the cavities of the substrate (Wenzel state), and the friction will be high. Farther downstream, in turn, the pressure is low and the liquid is more likely to adopt the Cassie state, which results in a lower friction. 
At negative pressures, $p<p_\textrm{coex}$, one might expect either again an increase of friction because the liquid-vapor interface in the Cassie state is curved (bubble mattress)~\cite{Steinberger2007} or a decrease of friction because the vapor layer at the solid is formed and acts as a perfect lubricant~\cite{OV_NB_NC_1995,CCB_JLB_LB_EC_2003}.

\begin{figure}[ht]
    \centering
    \subfloat[]{\label{fig:s2d1_slip_frict_vs_n}\includegraphics[height=5.4cm]{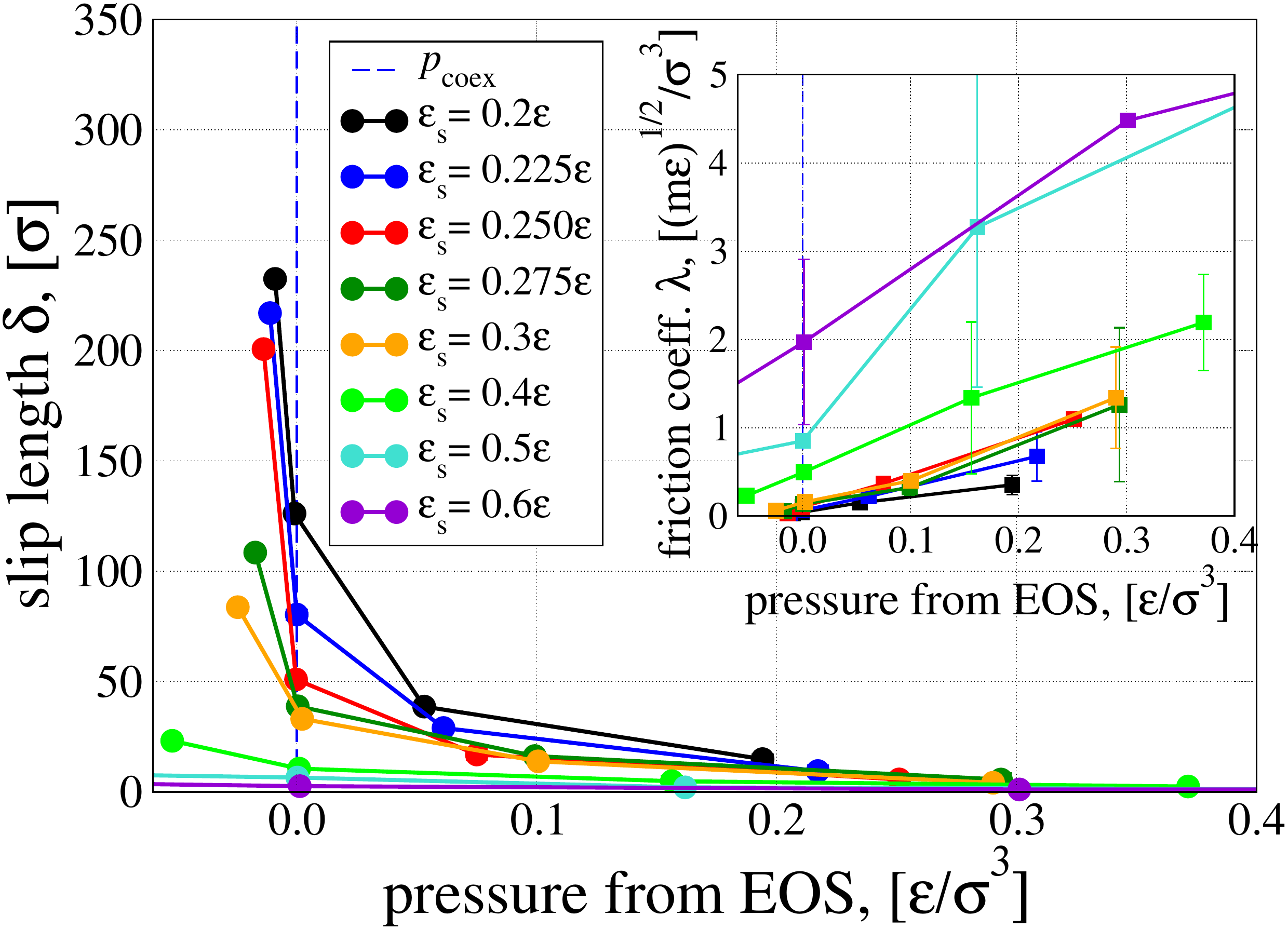}}\vspace*{0.4cm}
    
    \subfloat[]{\label{fig:s7d4_slip_frict_vs_n}\includegraphics[height=5.4cm]{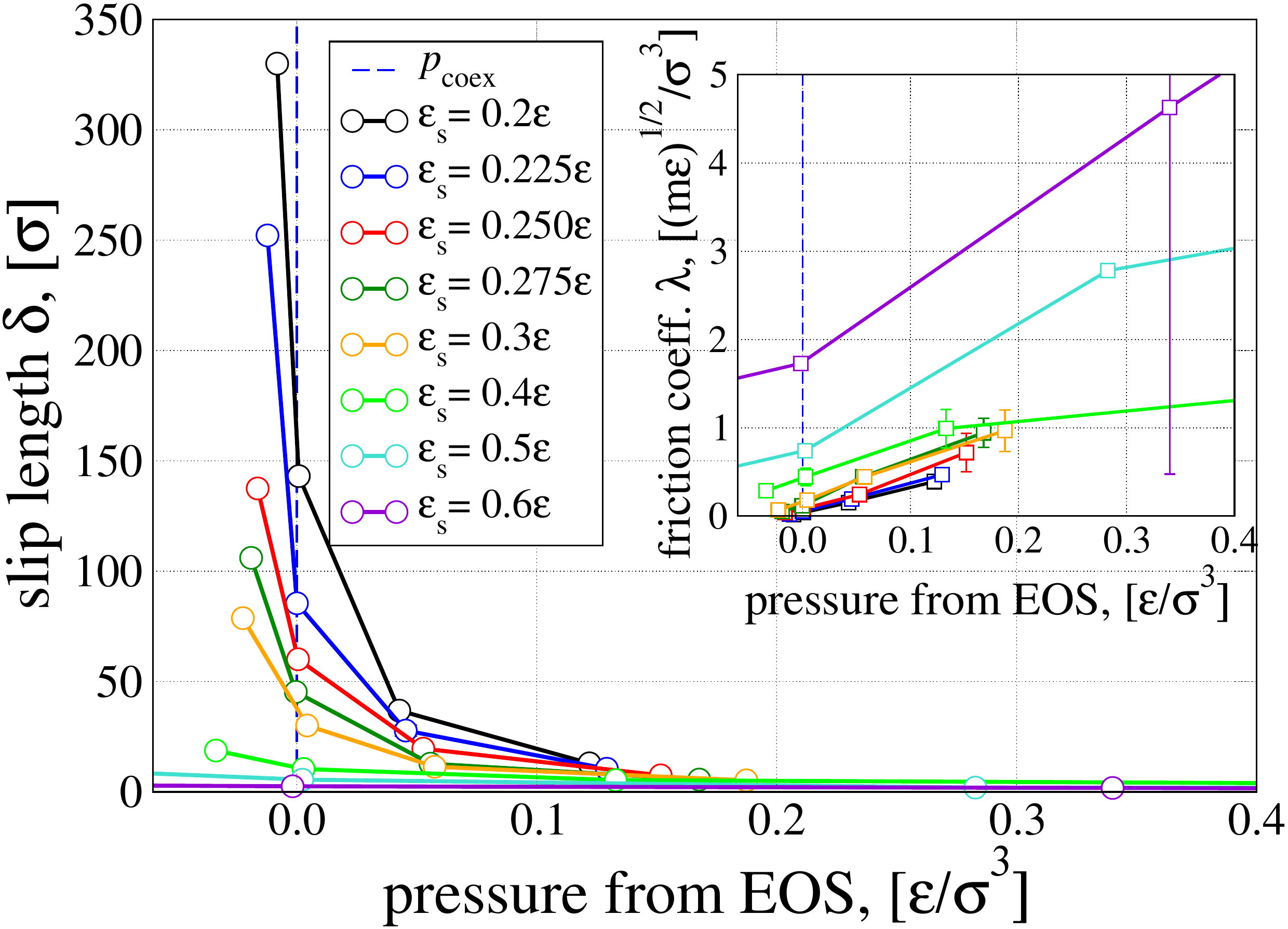}}
    \caption{(Color online) Slip length, $\delta$, as a function of the pressure measured from EOS of the channel confined by finely (a) or coarsely (b) corrugated substrates, respectively. The insets present the friction coefficient, $\lambda$, and the vertical line marks $p_{\rm coex}$. The widths of the channels $L_{z}$ at fixed $\epsilon_\textrm{s}$ and corrugation $r_i$ correspond to the widths $L_z^\textrm{coex}\,(\epsilon_\textrm{s},\,r_i,\,N_0)$, calculated in Sec.~\ref{sec:norm-press}.}
    \label{fig:var_press_slip_frict_vs_n}
\end{figure}

In order to study how the slip length depends on pressure, we generate Poiseuille-like flow via a body force and vary the number of polymers inside the slit-pore. We measure the density at the center of the channel and use the bulk equation of state (EOS) to determine the pressure in Fig.~\ref{fig:var_press_slip_frict_vs_n}. This pressure corresponds to the normal pressure inside the channel.

The simulation results in Fig.~\ref{fig:var_press_slip_frict_vs_n} demonstrate that the slip length sensitively depend on pressure. If the pressure is high, the slip length, $\delta$, is microscopically small and the friction coefficient $\lambda$ is very high. In this case we observe a parabolic profile with an effective no-slip boundary condition. Upon reducing the pressure, the slip length increases. When approaching the coexistence pressure, the slip length increases for weakly attractive substrates because the grooves are gradually emptied. At $p_{\rm coex}$ the liquid adopts the Cassie state. For stronger attractions, $\epsilon_\textrm{s}$ the liquid remains in the Wenzel state even when we approach $p_{\rm coex}$. 
The continuous transition between Cassie and Wenzel states at $p_{\rm coex}$ as a function of $\epsilon_\textrm{s}$ has been studied in Sec.~\ref{IIb}. For $p<p_{\rm coex}$ very large slip lengths can be achieved because a thin vapor layer is formed at the substrate that acts as lubricant. In this case, we observe plug flow because the slip lengths exceeds our system size, $L_{z}$ by far, and the friction is very small. Such a lubrication layer forms more readily for small $\epsilon_\textrm{s}$. 

\begin{figure}[ht]
    \subfloat[]{\label{fig:s2d1_b_pos_vs_eps}\includegraphics[height=3cm]{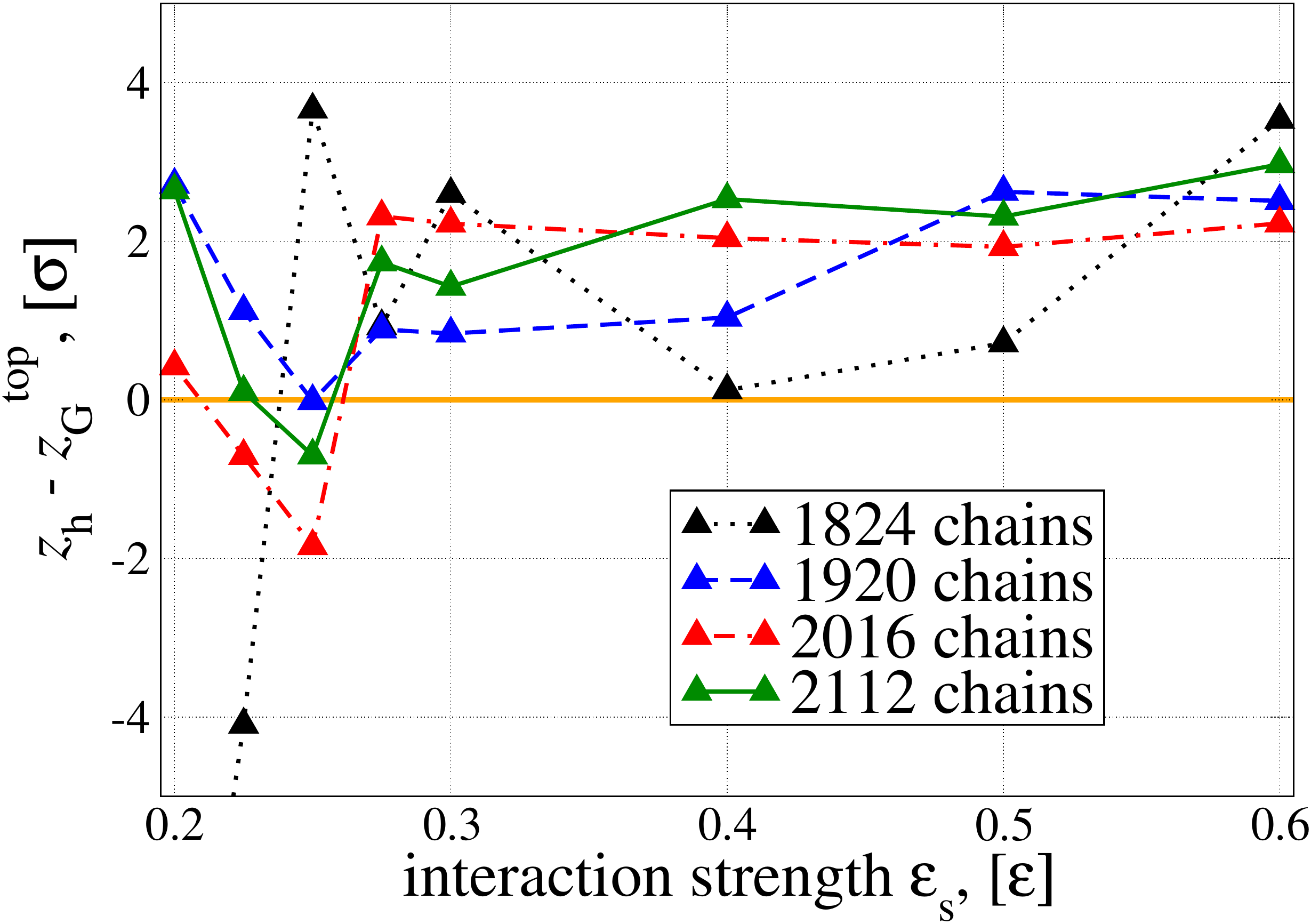}}\hspace*{0.1cm}   
    \subfloat[]{\label{fig:s7d4_b_pos_vs_eps}\includegraphics[height=3cm]{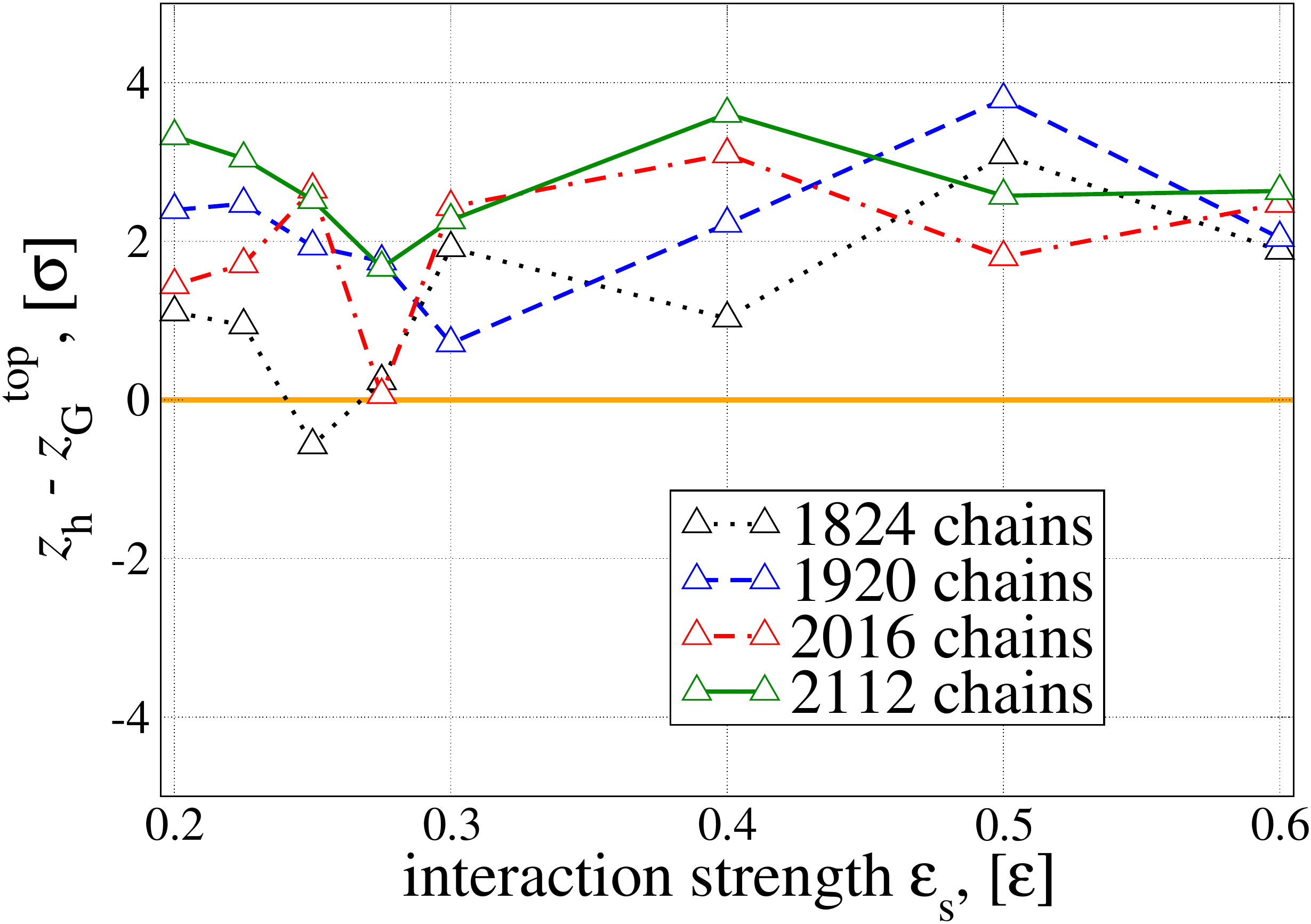}}
    \caption{(Color online) Distance of the position, $z_\textrm{h}$, at which Navier's boundary condition~\ref{eq:navier} is to be applied, to the innermost substrate atom $z_\textrm{G}^\textrm{top}$ for channels with fine (a) and rough (b) corrugations, respectively. Positive values, $z_\textrm{h} - z_\textrm{G}^\textrm{top} > 0$, indicate that the effective boundary position is closer to the center of the channel than any particle of the wall. Different line types correspond to different number of polymers in a channel at fixed $L_{z}=L_z^\textrm{coex}\,(\epsilon_\textrm{s},\,r_i,\,N_0)$, giving rise to different pressures.}
    \label{fig:var_press_b_pos_vs_eps}
\end{figure}

The dependence of the hydrodynamic boundary position, $z_\textrm{h}$, on pressure is shown in Fig.~\ref{fig:var_press_b_pos_vs_eps}. As in the case of flows at coexistence pressure, discussed in Sec.~\ref{sec:slip-frict}, for most parameters the hydrodynamic boundary is found inside the liquid above the top of the ridges between the grooves. The significant change of hydrodynamic position, $z_\textrm{h}$ at low pressures and at finely corrugated substrate, which goes along with a large statistical uncertainty, may be explained by the formation of a lubricant layer of vapor at the solid boundary and the concomitant, large increase of slippage (plug flow).

Thus slippage and friction greatly depend on the pressure inside the channel, and the pressure dependence may significantly affect the flow profile across the channel and the pressure drop along the channel. Using Navier's hydrodynamic boundary condition, we compute how the flow rate, $Q$, depends on the slip length, $\delta$, and half of the effective channel width, $z_\textrm{h}$

\begin{equation}
 Q = 2 \int^{L_y}_0 dy \int_{0}^{z_\textrm{h}} v_x^\textrm{P}(z) dz = - \frac{2 L_y z_\textrm{h}^3}{3 \eta} \frac{{\rm d}p}{{\rm d}x} \Big[3\frac{\delta(p)}{z_\textrm{h}} + 1\Big],
 \label{eq:flow_rate}
\end{equation}
where $ \frac{{\rm d}p}{{\rm d}z}$ is the local drop of pressure along the channel. Mass conservation requires that $Q$ be constant along the channel and the pressure profile along the channel is dictated by
\begin{equation}
\frac{{\rm d}p}{{\rm d}x} \sim - \frac{1}{3\frac{\delta(p(x))}{z_\textrm{h}}+1}
\end{equation}
If the slip length is constant along the channel, the pressure decreases {linearly} with the coordinate, $x$, along the channel. For the corrugated substrate, however, $\delta$ increases with $p$ and, consequentially, the pressure decreases rapidly upstream and more slowly downstream. In the case of negative pressures at the end of the channel (suction) and small $\epsilon_\textrm{s}$, we expect plug flow with vanishingly small pressure drop.

%
%
\section{Discussion and Outlook}
\label{sec:discout}
%
%

In this paper we studied the flows of a polymer liquid in channels with patterned walls by non-equilibrium Molecular Dynamics simulations and compared the results to the channels with flat walls. First, we measured the normal component of the pressure in the bulk at different widths of the channels of different topography and compared the slippage and the friction at the coexistence pressure. 

Significant differences between flat and corrugated substrates were found for liquids in Cassie state. Owing to the friction at the edges of the grooves, however, the friction coefficient $\lambda$ does not scale like the ratio of area covered by the grooves, $\varphi$, to the projected area. For microscopically corrugated substrates, a correction associated with the fiction at the edges of the corrugations is suggested, c.f.~Eq.~\ref{eq:fric_pinned}. Contrary to the expectation, the hydrodynamic properties of the liquid in Wenzel state are not significantly influenced by the substrate corrugation, $r$. For the parameters of our simulation, the difference with respect to a flat substrate is rather controlled by the area fraction $\varphi$, but not the size of the grooves. We show that for substrates with small depth of corrugations there is no sharp phase transition between Cassie and Wenzel states but only a rather gradual crossover.
These observations indicate that the macroscopic concepts cannot be straightforwardly extrapolated down to substrate topographies with dimensions that correspond to tens of fluid particles.

Moreover, we illustrated that one configuration of the flow is insufficient for a measurement of the hydrodynamic boundary condition, i.e., a  boundary condition to the macroscopic hydrodynamic continuum equations that describes the substrate properties independently from the type and strengths of the flow. Using Couette and Poiseuille flow, we extracted the hydrodynamic boundary position, $z_\textrm{h}$,
and slip length, $\delta$, that characterize the Navier slip condition. In applications to superhydrophobic substrates of complex topography the position of the hydrodynamic boundary is not intuitive. While the equivalent no-slip plane, which depends on the type of flow, might be located between the top and bottom of the grooves~\cite{CK_JH_OV_2010}, we find that the position of the hydrodynamic boundary, $z_\textrm{h}$, was almost always located above the top of the roughness. 

Finally, we modeled the propagation of the liquid through a long microfluidic channel with patterned walls accounting for the pressure dependence of the liquid morphology and the concomitant friction. We observed that by using a non-wetting liquid (contact angle on a flat substrate $\theta_\textrm{E} > 150 \text{\textdegree}$) the rate of the frictional dissipation is significantly reduced on the whole length of a long channel in comparison to more wettable substrates. Additionally, we found that only for a liquid at low pressure {($p < p_\textrm{coex}$)} the microscopic corrugation affects the hydrodynamic properties, whereas for higher pressures it does not exert a pronounced influence onto the slippage and friction.

%
\section{Acknowledgment}
\label{sec:acknow}

The authors thank F.~L{\'e}onforte, A.~Galuschko, A.O.~Parry for fruitful and inspiring discussions and the GWDG computing center at G\"ottingen, J\"ulich Supercomputing Centre (JSC) and Computing Centre in Hannover (HLRN) for the computational resources. This work was supported by the European Union under grant PITN-GA-2008-214919 (MULTIFLOW).
%

\bibliography{TM_slip_2012}

\providecommand*{\mcitethebibliography}{\thebibliography}
\csname @ifundefined\endcsname{endmcitethebibliography}
{\let\endmcitethebibliography\endthebibliography}{}
\begin{mcitethebibliography}{48}
\providecommand*{\natexlab}[1]{#1}
\providecommand*{\mciteSetBstSublistMode}[1]{}
\providecommand*{\mciteSetBstMaxWidthForm}[2]{}
\providecommand*{\mciteBstWouldAddEndPuncttrue}
  {\def\EndOfBibitem{\unskip.}}
\providecommand*{\mciteBstWouldAddEndPunctfalse}
  {\let\EndOfBibitem\relax}
\providecommand*{\mciteSetBstMidEndSepPunct}[3]{}
\providecommand*{\mciteSetBstSublistLabelBeginEnd}[3]{}
\providecommand*{\EndOfBibitem}{}
\mciteSetBstSublistMode{f}
\mciteSetBstMaxWidthForm{subitem}
{(\emph{\alph{mcitesubitemcount}})}
\mciteSetBstSublistLabelBeginEnd{\mcitemaxwidthsubitemform\space}
{\relax}{\relax}

\bibitem[Ou \emph{et~al.}(2004)Ou, Perot, and Rothstein]{JO_BP_JR_2004}
J.~Ou, B.~Perot and J.~P. Rothstein, \emph{Physics of Fluids}, 2004,
  \textbf{16}, 4635--4643\relax
\mciteBstWouldAddEndPuncttrue
\mciteSetBstMidEndSepPunct{\mcitedefaultmidpunct}
{\mcitedefaultendpunct}{\mcitedefaultseppunct}\relax
\EndOfBibitem
\bibitem[Rothstein(2010)]{Rothstein_2010}
J.~P. Rothstein, \emph{Annual Review of Fluid Mechanics}, 2010, \textbf{42},
  89--109\relax
\mciteBstWouldAddEndPuncttrue
\mciteSetBstMidEndSepPunct{\mcitedefaultmidpunct}
{\mcitedefaultendpunct}{\mcitedefaultseppunct}\relax
\EndOfBibitem
\bibitem[{Queralt-Mart{\'{\i}}n} \emph{et~al.}(2011){Queralt-Mart{\'{\i}}n},
  {Pradas}, {Rodr{\'{\i}}guez-Trujillo}, {Arundell}, {Corvera Poir{\'e}}, and
  {Hern{\'a}ndez-Machado}]{MQM_MP_RRT_2011}
M.~{Queralt-Mart{\'{\i}}n}, M.~{Pradas}, R.~{Rodr{\'{\i}}guez-Trujillo},
  M.~{Arundell}, E.~{Corvera Poir{\'e}} and A.~{Hern{\'a}ndez-Machado},
  \emph{Phys. Rev. Lett.}, 2011, \textbf{106}, 194501\relax
\mciteBstWouldAddEndPuncttrue
\mciteSetBstMidEndSepPunct{\mcitedefaultmidpunct}
{\mcitedefaultendpunct}{\mcitedefaultseppunct}\relax
\EndOfBibitem
\bibitem[Cottin-Bizonne \emph{et~al.}(2003)Cottin-Bizonne, Barrat, Bocquet, and
  Charlaix]{CCB_JLB_LB_EC_2003}
C.~Cottin-Bizonne, J.-L. Barrat, L.~Bocquet and E.~Charlaix, \emph{Nat.
  Mater.}, 2003, \textbf{2}, 237--240\relax
\mciteBstWouldAddEndPuncttrue
\mciteSetBstMidEndSepPunct{\mcitedefaultmidpunct}
{\mcitedefaultendpunct}{\mcitedefaultseppunct}\relax
\EndOfBibitem
\bibitem[Galea and Attard(2004)]{TG_PA_2004}
T.~M. Galea and P.~Attard, \emph{Langmuir}, 2004, \textbf{20}, 3477--3482\relax
\mciteBstWouldAddEndPuncttrue
\mciteSetBstMidEndSepPunct{\mcitedefaultmidpunct}
{\mcitedefaultendpunct}{\mcitedefaultseppunct}\relax
\EndOfBibitem
\bibitem[Kim and Darve(2006)]{DK_ED_2006}
D.~Kim and E.~Darve, \emph{Phys. Rev. E}, 2006, \textbf{73}, 051203\relax
\mciteBstWouldAddEndPuncttrue
\mciteSetBstMidEndSepPunct{\mcitedefaultmidpunct}
{\mcitedefaultendpunct}{\mcitedefaultseppunct}\relax
\EndOfBibitem
\bibitem[Wenzel(1936)]{Wenzel_1936}
R.~N. Wenzel, \emph{Industrial \& Engineering Chemistry}, 1936, \textbf{28},
  988--994\relax
\mciteBstWouldAddEndPuncttrue
\mciteSetBstMidEndSepPunct{\mcitedefaultmidpunct}
{\mcitedefaultendpunct}{\mcitedefaultseppunct}\relax
\EndOfBibitem
\bibitem[Cassie and Baxter(1944)]{AC_SB_1944}
A.~B.~D. Cassie and S.~Baxter, \emph{Trans. Faraday Soc.}, 1944, \textbf{40},
  546--551\relax
\mciteBstWouldAddEndPuncttrue
\mciteSetBstMidEndSepPunct{\mcitedefaultmidpunct}
{\mcitedefaultendpunct}{\mcitedefaultseppunct}\relax
\EndOfBibitem
\bibitem[Young(1805)]{Young}
T.~Young, \emph{Philos.\ Trans.\ R.\ Soc.\ London}, 1805, \textbf{95},
  65--87\relax
\mciteBstWouldAddEndPuncttrue
\mciteSetBstMidEndSepPunct{\mcitedefaultmidpunct}
{\mcitedefaultendpunct}{\mcitedefaultseppunct}\relax
\EndOfBibitem
\bibitem[Bico \emph{et~al.}(1999)Bico, Marzolin, and Qu\'{e}r\'{e}]{Bico1999}
J.~Bico, C.~Marzolin and D.~Qu\'{e}r\'{e}, \emph{Europhys. Lett.}, 1999,
  \textbf{47}, 220\relax
\mciteBstWouldAddEndPuncttrue
\mciteSetBstMidEndSepPunct{\mcitedefaultmidpunct}
{\mcitedefaultendpunct}{\mcitedefaultseppunct}\relax
\EndOfBibitem
\bibitem[Qu\'{e}r\'{e} \emph{et~al.}(2003)Qu\'{e}r\'{e}, Lafuma, and
  Bico]{Bico2003}
D.~Qu\'{e}r\'{e}, A.~Lafuma and J.~Bico, \emph{Nanotechnology}, 2003,
  \textbf{14}, 1109\relax
\mciteBstWouldAddEndPuncttrue
\mciteSetBstMidEndSepPunct{\mcitedefaultmidpunct}
{\mcitedefaultendpunct}{\mcitedefaultseppunct}\relax
\EndOfBibitem
\bibitem[Navier(1823)]{Navier_1823}
C.~L. M.~H. Navier, \emph{Mem. Acad. Sci. Inst. Fr.}, 1823, \textbf{6},
  432--6\relax
\mciteBstWouldAddEndPuncttrue
\mciteSetBstMidEndSepPunct{\mcitedefaultmidpunct}
{\mcitedefaultendpunct}{\mcitedefaultseppunct}\relax
\EndOfBibitem
\bibitem[Granick \emph{et~al.}(2003)Granick, Zhu, and Lee]{SG_YZ_HL_2003}
S.~Granick, Y.~Zhu and H.~Lee, \emph{Nat. Mater.}, 2003, \textbf{2},
  221--227\relax
\mciteBstWouldAddEndPuncttrue
\mciteSetBstMidEndSepPunct{\mcitedefaultmidpunct}
{\mcitedefaultendpunct}{\mcitedefaultseppunct}\relax
\EndOfBibitem
\bibitem[Cottin-Bizonne \emph{et~al.}(2004)Cottin-Bizonne, Barentin, Charlaix,
  Bocquet, and Barrat]{CCB_CB_EC_2004}
C.~Cottin-Bizonne, C.~Barentin, E.~Charlaix, L.~Bocquet and J.~Barrat,
  \emph{The European Physical Journal E: Soft Matter and Biological Physics},
  2004, \textbf{15}, 427--438\relax
\mciteBstWouldAddEndPuncttrue
\mciteSetBstMidEndSepPunct{\mcitedefaultmidpunct}
{\mcitedefaultendpunct}{\mcitedefaultseppunct}\relax
\EndOfBibitem
\bibitem[Kunert \emph{et~al.}(2010)Kunert, Harting, and
  Vinogradova]{CK_JH_OV_2010}
C.~Kunert, J.~Harting and O.~I. Vinogradova, \emph{Phys. Rev. Lett.}, 2010,
  \textbf{105}, 016001\relax
\mciteBstWouldAddEndPuncttrue
\mciteSetBstMidEndSepPunct{\mcitedefaultmidpunct}
{\mcitedefaultendpunct}{\mcitedefaultseppunct}\relax
\EndOfBibitem
\bibitem[Vinogradova and Belyaev(2011)]{OV_AB_2011}
O.~I. Vinogradova and A.~V. Belyaev, \emph{J. Phys.: Condens. Matter}, 2011,
  \textbf{23}, 184104\relax
\mciteBstWouldAddEndPuncttrue
\mciteSetBstMidEndSepPunct{\mcitedefaultmidpunct}
{\mcitedefaultendpunct}{\mcitedefaultseppunct}\relax
\EndOfBibitem
\bibitem[Giacomello \emph{et~al.}(2012)Giacomello, Chinappi, Meloni, and
  Casciola]{AG_MC_SM_2012}
A.~Giacomello, M.~Chinappi, S.~Meloni and C.~M. Casciola, \emph{Phys. Rev.
  Lett.}, 2012, \textbf{109}, 226102\relax
\mciteBstWouldAddEndPuncttrue
\mciteSetBstMidEndSepPunct{\mcitedefaultmidpunct}
{\mcitedefaultendpunct}{\mcitedefaultseppunct}\relax
\EndOfBibitem
\bibitem[M{\"u}ller and Mac{D}owell(2000)]{Muller00f}
M.~M{\"u}ller and L.~G. Mac{D}owell, \emph{Macromolecules}, 2000, \textbf{33},
  3902--3923\relax
\mciteBstWouldAddEndPuncttrue
\mciteSetBstMidEndSepPunct{\mcitedefaultmidpunct}
{\mcitedefaultendpunct}{\mcitedefaultseppunct}\relax
\EndOfBibitem
\bibitem[M{\"u}ller and Mac{D}owell(2003)]{Muller03c}
M.~M{\"u}ller and L.~G. Mac{D}owell, \emph{J. Phys.: Condens. Matter}, 2003,
  \textbf{15}, R609--R653\relax
\mciteBstWouldAddEndPuncttrue
\mciteSetBstMidEndSepPunct{\mcitedefaultmidpunct}
{\mcitedefaultendpunct}{\mcitedefaultseppunct}\relax
\EndOfBibitem
\bibitem[R.~B.~Bird(1977)]{Bird77}
O.~H. R.~B.~Bird, R.C.~Armstrong, \emph{Dynamics of Polymeric Liquids}, Wiley,
  New York, 1977, vol. 1, 2\relax
\mciteBstWouldAddEndPuncttrue
\mciteSetBstMidEndSepPunct{\mcitedefaultmidpunct}
{\mcitedefaultendpunct}{\mcitedefaultseppunct}\relax
\EndOfBibitem
\bibitem[Kremer and Grest(1990)]{KK_GG_1990}
K.~Kremer and G.~S. Grest, \emph{J. Chem. Phys.}, 1990, \textbf{92},
  5057--5086\relax
\mciteBstWouldAddEndPuncttrue
\mciteSetBstMidEndSepPunct{\mcitedefaultmidpunct}
{\mcitedefaultendpunct}{\mcitedefaultseppunct}\relax
\EndOfBibitem
\bibitem[Hoogerbrugge and Koelman(1992)]{PH_JK_1992}
P.~J. Hoogerbrugge and J.~M. V.~A. Koelman, \emph{Europhys. Lett.}, 1992,
  \textbf{19}, 155\relax
\mciteBstWouldAddEndPuncttrue
\mciteSetBstMidEndSepPunct{\mcitedefaultmidpunct}
{\mcitedefaultendpunct}{\mcitedefaultseppunct}\relax
\EndOfBibitem
\bibitem[Espa\~{n}ol and Warren(1995)]{PE_PW_1995}
P.~Espa\~{n}ol and P.~Warren, \emph{Europhys. Lett.}, 1995, \textbf{30},
  191\relax
\mciteBstWouldAddEndPuncttrue
\mciteSetBstMidEndSepPunct{\mcitedefaultmidpunct}
{\mcitedefaultendpunct}{\mcitedefaultseppunct}\relax
\EndOfBibitem
\bibitem[Servantie and M\"{u}ller(2008)]{JS_MM_2008}
J.~Servantie and M.~M\"{u}ller, \emph{J. Chem. Phys.}, 2008, \textbf{128},
  014709\relax
\mciteBstWouldAddEndPuncttrue
\mciteSetBstMidEndSepPunct{\mcitedefaultmidpunct}
{\mcitedefaultendpunct}{\mcitedefaultseppunct}\relax
\EndOfBibitem
\bibitem[Tretyakov \emph{et~al.}(2012)Tretyakov, M\"{u}ller, Todorova, and
  Thiele]{NT_MM_DT_UT_12}
N.~Tretyakov, M.~M\"{u}ller, D.~Todorova and U.~Thiele, \emph{J. Chem. Phys.},
  2012\relax
\mciteBstWouldAddEndPuncttrue
\mciteSetBstMidEndSepPunct{\mcitedefaultmidpunct}
{\mcitedefaultendpunct}{\mcitedefaultseppunct}\relax
\EndOfBibitem
\bibitem[Parry(2012)]{Andy}
A.~O. Parry, \emph{private communication}, 2012\relax
\mciteBstWouldAddEndPuncttrue
\mciteSetBstMidEndSepPunct{\mcitedefaultmidpunct}
{\mcitedefaultendpunct}{\mcitedefaultseppunct}\relax
\EndOfBibitem
\bibitem[Parry \emph{et~al.}(2007)Parry, Rascon, Wilding, and
  Evans]{AP_CR_NW_2007}
A.~O. Parry, C.~Rascon, N.~B. Wilding and R.~Evans, \emph{Phys. Rev. Lett.},
  2007, \textbf{98}, 226101\relax
\mciteBstWouldAddEndPuncttrue
\mciteSetBstMidEndSepPunct{\mcitedefaultmidpunct}
{\mcitedefaultendpunct}{\mcitedefaultseppunct}\relax
\EndOfBibitem
\bibitem[Roth and Parry(2011)]{RR_AP_2011}
R.~Roth and A.~O. Parry, \emph{Mol. Phys.}, 2011, \textbf{109},
  1159--1167\relax
\mciteBstWouldAddEndPuncttrue
\mciteSetBstMidEndSepPunct{\mcitedefaultmidpunct}
{\mcitedefaultendpunct}{\mcitedefaultseppunct}\relax
\EndOfBibitem
\bibitem[Allen and Tildesley(1989)]{AllenTild89}
M.~Allen and D.~Tildesley, \emph{Computer simulation of liquids}, Clarendon
  Press, 1989\relax
\mciteBstWouldAddEndPuncttrue
\mciteSetBstMidEndSepPunct{\mcitedefaultmidpunct}
{\mcitedefaultendpunct}{\mcitedefaultseppunct}\relax
\EndOfBibitem
\bibitem[Frenkel and Smit(2002)]{FrenkSmit02}
D.~Frenkel and B.~Smit, \emph{Understanding molecular simulation: from
  algorithms to applications}, Academic Press, 2002\relax
\mciteBstWouldAddEndPuncttrue
\mciteSetBstMidEndSepPunct{\mcitedefaultmidpunct}
{\mcitedefaultendpunct}{\mcitedefaultseppunct}\relax
\EndOfBibitem
\bibitem[Irving and Kirkwood(1950)]{JI_JK_1950}
J.~H. Irving and J.~G. Kirkwood, \emph{J. Chem. Phys.}, 1950, \textbf{18},
  817--829\relax
\mciteBstWouldAddEndPuncttrue
\mciteSetBstMidEndSepPunct{\mcitedefaultmidpunct}
{\mcitedefaultendpunct}{\mcitedefaultseppunct}\relax
\EndOfBibitem
\bibitem[Barrat and Bocquet(1999)]{JLB_LB_1999}
J.-L. Barrat and L.~Bocquet, \emph{Phys. Rev. Lett.}, 1999, \textbf{82},
  4671--4674\relax
\mciteBstWouldAddEndPuncttrue
\mciteSetBstMidEndSepPunct{\mcitedefaultmidpunct}
{\mcitedefaultendpunct}{\mcitedefaultseppunct}\relax
\EndOfBibitem
\bibitem[Barrat and Bocquet(1999)]{JLB_LB_1999_FD}
J.-L. Barrat and L.~Bocquet, \emph{Faraday Discuss.}, 1999, \textbf{112},
  119--128\relax
\mciteBstWouldAddEndPuncttrue
\mciteSetBstMidEndSepPunct{\mcitedefaultmidpunct}
{\mcitedefaultendpunct}{\mcitedefaultseppunct}\relax
\EndOfBibitem
\bibitem[M\"{u}ller and Pastorino(2008)]{MM_CP_2008}
M.~M\"{u}ller and C.~Pastorino, \emph{Europhys. Lett.}, 2008, \textbf{81},
  28002\relax
\mciteBstWouldAddEndPuncttrue
\mciteSetBstMidEndSepPunct{\mcitedefaultmidpunct}
{\mcitedefaultendpunct}{\mcitedefaultseppunct}\relax
\EndOfBibitem
\bibitem[Smiatek \emph{et~al.}(2008)Smiatek, Allen, and Schmid]{JS_MA_FS_2008}
J.~Smiatek, M.~Allen and F.~Schmid, \emph{The European Physical Journal E: Soft
  Matter and Biological Physics}, 2008, \textbf{26}, 115--122\relax
\mciteBstWouldAddEndPuncttrue
\mciteSetBstMidEndSepPunct{\mcitedefaultmidpunct}
{\mcitedefaultendpunct}{\mcitedefaultseppunct}\relax
\EndOfBibitem
\bibitem[Bocquet and Barrat(2007)]{LB_JLB_2007}
L.~Bocquet and J.-L. Barrat, \emph{Soft Matter}, 2007, \textbf{3},
  685--693\relax
\mciteBstWouldAddEndPuncttrue
\mciteSetBstMidEndSepPunct{\mcitedefaultmidpunct}
{\mcitedefaultendpunct}{\mcitedefaultseppunct}\relax
\EndOfBibitem
\bibitem[M\"{u}ller \emph{et~al.}(2008)M\"{u}ller, Pastorino, and
  Servantie]{MM_CP_JS_2008}
M.~M\"{u}ller, C.~Pastorino and J.~Servantie, \emph{J. Phys.: Condens. Matter},
  2008, \textbf{20}, 494225\relax
\mciteBstWouldAddEndPuncttrue
\mciteSetBstMidEndSepPunct{\mcitedefaultmidpunct}
{\mcitedefaultendpunct}{\mcitedefaultseppunct}\relax
\EndOfBibitem
\bibitem[B\"{a}umchen and Jacobs(2010)]{OB_KJ_2010}
O.~B\"{a}umchen and K.~Jacobs, \emph{J. Phys.: Condens. Matter}, 2010,
  \textbf{22}, 033102\relax
\mciteBstWouldAddEndPuncttrue
\mciteSetBstMidEndSepPunct{\mcitedefaultmidpunct}
{\mcitedefaultendpunct}{\mcitedefaultseppunct}\relax
\EndOfBibitem
\bibitem[L{\'e}onforte \emph{et~al.}(2011)L{\'e}onforte, Servantie, Pastorino,
  and M{\"u}ller]{FL_JS_CP_MM_2011}
F.~L{\'e}onforte, J.~Servantie, C.~Pastorino and M.~M{\"u}ller, \emph{J. Phys.:
  Condens. Matter}, 2011, \textbf{23}, 184105\relax
\mciteBstWouldAddEndPuncttrue
\mciteSetBstMidEndSepPunct{\mcitedefaultmidpunct}
{\mcitedefaultendpunct}{\mcitedefaultseppunct}\relax
\EndOfBibitem
\bibitem[Cottin-Bizonne \emph{et~al.}(2002)Cottin-Bizonne, Jurine, Baudry,
  Crassous, Restagno, and Charlaix]{CCB_SJ_JB_2002}
C.~Cottin-Bizonne, S.~Jurine, J.~Baudry, J.~Crassous, F.~Restagno and
  E.~Charlaix, \emph{The European Physical Journal E: Soft Matter and
  Biological Physics}, 2002, \textbf{9}, 47--53\relax
\mciteBstWouldAddEndPuncttrue
\mciteSetBstMidEndSepPunct{\mcitedefaultmidpunct}
{\mcitedefaultendpunct}{\mcitedefaultseppunct}\relax
\EndOfBibitem
\bibitem[Lee \emph{et~al.}(2008)Lee, Choi, and Kim]{CL_CHC_CLK_2008}
C.~Lee, C.-H. Choi and C.-J.~C. Kim, \emph{Phys. Rev. Lett.}, 2008,
  \textbf{101}, 064501\relax
\mciteBstWouldAddEndPuncttrue
\mciteSetBstMidEndSepPunct{\mcitedefaultmidpunct}
{\mcitedefaultendpunct}{\mcitedefaultseppunct}\relax
\EndOfBibitem
\bibitem[Schmieschek \emph{et~al.}(2012)Schmieschek, Belyaev, Harting, and
  Vinogradova]{SS_AB_JH_OV_2012}
S.~Schmieschek, A.~V. Belyaev, J.~Harting and O.~I. Vinogradova, \emph{Phys.
  Rev. E}, 2012, \textbf{85}, 016324\relax
\mciteBstWouldAddEndPuncttrue
\mciteSetBstMidEndSepPunct{\mcitedefaultmidpunct}
{\mcitedefaultendpunct}{\mcitedefaultseppunct}\relax
\EndOfBibitem
\bibitem[Qian \emph{et~al.}(2005)Qian, Wang, and Sheng]{TQ_XPW_PS_2005}
T.~Qian, X.-P. Wang and P.~Sheng, \emph{Phys. Rev. E}, 2005, \textbf{72},
  022501\relax
\mciteBstWouldAddEndPuncttrue
\mciteSetBstMidEndSepPunct{\mcitedefaultmidpunct}
{\mcitedefaultendpunct}{\mcitedefaultseppunct}\relax
\EndOfBibitem
\bibitem[Priezjev(2011)]{Priezjev_2011}
N.~V. Priezjev, \emph{J. Chem. Phys.}, 2011, \textbf{135}, 204704\relax
\mciteBstWouldAddEndPuncttrue
\mciteSetBstMidEndSepPunct{\mcitedefaultmidpunct}
{\mcitedefaultendpunct}{\mcitedefaultseppunct}\relax
\EndOfBibitem
\bibitem[Servantie and M{\"u}ller(2008)]{Servantie08b}
J.~Servantie and M.~M{\"u}ller, \emph{Phys. Rev. Lett.}, 2008, \textbf{101},
  4989--5001\relax
\mciteBstWouldAddEndPuncttrue
\mciteSetBstMidEndSepPunct{\mcitedefaultmidpunct}
{\mcitedefaultendpunct}{\mcitedefaultseppunct}\relax
\EndOfBibitem
\bibitem[Soong \emph{et~al.}(2007)Soong, Yen, and Tzeng]{CS_TY_PT_2007}
C.~Y. Soong, T.~H. Yen and P.~Y. Tzeng, \emph{Phys. Rev. E}, 2007, \textbf{76},
  036303\relax
\mciteBstWouldAddEndPuncttrue
\mciteSetBstMidEndSepPunct{\mcitedefaultmidpunct}
{\mcitedefaultendpunct}{\mcitedefaultseppunct}\relax
\EndOfBibitem
\bibitem[Steinberger \emph{et~al.}(2007)Steinberger, Cottin-Bizonne, Kleimann,
  and Charlaix]{Steinberger2007}
A.~Steinberger, C.~Cottin-Bizonne, P.~Kleimann and E.~Charlaix, \emph{Nat.
  Mater.}, 2007, \textbf{6}, 665--668\relax
\mciteBstWouldAddEndPuncttrue
\mciteSetBstMidEndSepPunct{\mcitedefaultmidpunct}
{\mcitedefaultendpunct}{\mcitedefaultseppunct}\relax
\EndOfBibitem
\bibitem[Vinogradova \emph{et~al.}(1995)Vinogradova, Bunkin, Churaev, Kiseleva,
  Lobeyev, and Ninham]{OV_NB_NC_1995}
O.~Vinogradova, N.~Bunkin, N.~Churaev, O.~Kiseleva, A.~Lobeyev and B.~Ninham,
  \emph{Journal of Colloid and Interface Science}, 1995, \textbf{173}, 443 --
  447\relax
\mciteBstWouldAddEndPuncttrue
\mciteSetBstMidEndSepPunct{\mcitedefaultmidpunct}
{\mcitedefaultendpunct}{\mcitedefaultseppunct}\relax
\EndOfBibitem
\end{mcitethebibliography}

\bibliographystyle{rsc} 
\end{document}